\documentclass[sigconf,sigconf]{acmart}

\usepackage{ifthen}
\newcommand{\todo}[1]{\textcolor{teal}{[TODO: #1]}}
\newcommand{\obs}[1]{\textcolor{red}{[?: #1]}}
\newcommand{\new}[1]{{#1}}

\newcommand{\bd}[1]{\textcolor{blue}{[BD: #1]}}
\newcommand{\jy}[1]{\textcolor{blue}{[Jun: #1]}}
\newcommand{\TurnCommentsOn}[1]{%
  \ifthenelse{\equal{#1}{0}}{
    \renewcommand{\obs}[1]{}
    \renewcommand{\bd}[1]{}
    \renewcommand{\jy}[1]{}
    }{}}
\newcommand{\TurnTodoOn}[1]{%
  \ifthenelse{\equal{#1}{0}}{
    \renewcommand{\todo}[1]{}}{}}

\TurnCommentsOn{1}
\TurnTodoOn{0}

\copyrightyear{2023} 
\acmYear{2023} 
\setcopyright{acmlicensed}\acmConference[CHI '23]{Proceedings of the 2023 CHI Conference on Human Factors in Computing Systems}{April 23--28, 2023}{Hamburg, Germany}
\acmBooktitle{Proceedings of the 2023 CHI Conference on Human Factors in Computing Systems (CHI '23), April 23--28, 2023, Hamburg, Germany}
\acmPrice{15.00}
\acmDOI{10.1145/3544548.3581431}
\acmISBN{978-1-4503-9421-5/23/04}

\acmConference[CHI '23]{note}{April 23--28, 2023}{Hamburg, Germany}

\begin{document}

\title{Interface Design for Crowdsourcing Hierarchical Multi-Label Text Annotations}

\author{Rickard Stureborg}
\affiliation{%
  \institution{Duke University}
  \city{Durham}\state{NC}\country{USA}}
\email{rickard.stureborg@duke.edu}

\author{Bhuwan Dhingra}
\affiliation{%
  \institution{Duke University}
  \city{Durham}\state{NC}\country{USA}}
\email{bdhingra@cs.duke.edu}

\author{Jun Yang}
\affiliation{%
  \institution{Duke University}
  \city{Durham}\state{NC}\country{USA}}
\email{junyang@cs.duke.edu}

\renewcommand{\shortauthors}{Stureborg et al.}

\begin{abstract}

    Human data labeling is an important and expensive task at the heart of supervised learning systems.
    Hierarchies help humans understand and organize concepts.
    We ask whether and how concept hierarchies can inform the design of annotation interfaces to improve labeling quality and efficiency.
    We study this question through annotation of vaccine misinformation, where the labeling task is difficult and highly subjective.
    We investigate 6 user interface designs for crowdsourcing hierarchical labels by collecting over 18,000 individual annotations.
    Under a fixed budget, integrating hierarchies into the design improves crowdsource workers' F1 scores. We attribute this to
        (1) Grouping similar concepts, improving F1 scores by +0.16 over random groupings,
        (2) Strong relative performance on high-difficulty examples (relative F1 score difference of +0.40),
        and (3) Filtering out obvious negatives, increasing precision by +0.07. 
    Ultimately, labeling schemes integrating the hierarchy outperform those that do not --- achieving mean F1 of 0.70.

\end{abstract}



\maketitle

\section{Introduction}
\label{introduction}

To both build and evaluate machine learning systems, researchers often rely on human-labeled datasets 
    \cite{imagenet,rajpurkar-etal-2016-squad,multinli}.
Gathering this labeled data efficiently and at high quality is a well-studied problem when labels are 
    binary \cite{kay_kinetics_2017, yu_lsun_2016, celeb_faces} or 
    a flat list of choices \cite{swanson_snapshot_2015, lintott_galaxy_2008, deng_scalable_2014}, but
labels can often be grouped into other
structures as well, 
\new{such as species in a taxonomy \cite{inaturalist}}.

\begin{figure}[H]
   \vspace{0.6cm}
   \centering
   \includegraphics[width=.45\textwidth]{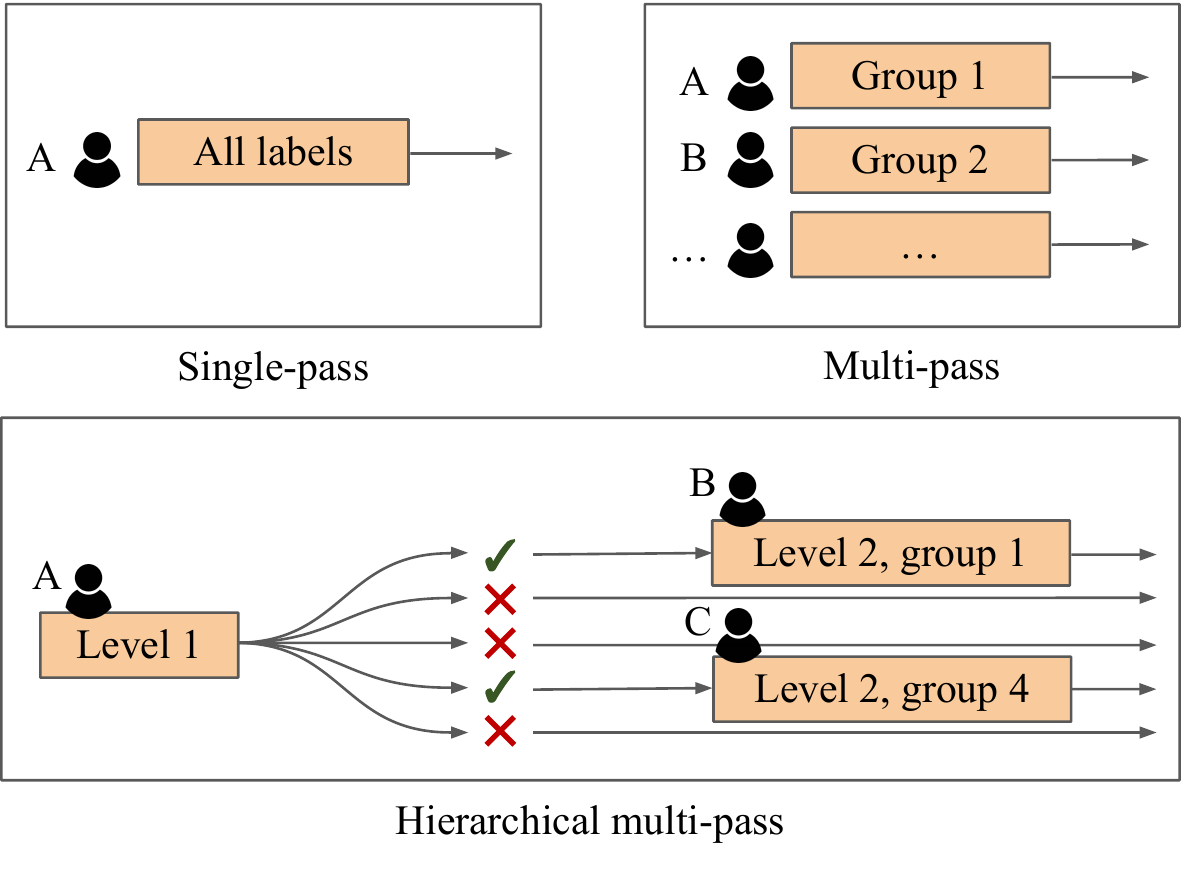}
   \caption{
    Pass-logic options when annotating a single passage. 
    Each orange box represents a single question asked to workers (denoted A, B, ...)
    and ``groups'' refers to partitioning the labels into smaller sets of labels.
    Workers perform best on mean F1 score when using hierarchical multi-pass schemes.
    }
 \Description{Three diagrams are shown describing single-pass, multi-pass, and hierarchical multi-pass routing logic. For single-pass, all set of labels are given to one worker. For multi-pass, the labels are partitioned into groups (1,2,...) and given to separate workers (A,B,...). For hierarchical multi-label, the top level labels are given to one worker, who's annotations determine whether or not the child labels will be given as a new group to annotators downstream. This example shows the case where the first worker labels TFFTF, and downstream there are two tasks set up for new workers to label the children of label 1 and label 4, respectively.}
   \label{fig:passlogic}
\end{figure}

Concept hierarchies (or taxonomies; ontologies) are used in many applications (\cite{acm_ccs, gemmeke_audio_2017, coan_climate_2021}) to describe concepts at a flexible granularity, and generally serve to help organize and structure both language and thought around a topic.
In certain situations, they may become a target for data labeling itself, where multiple hierarchically-structured class labels can be chosen for a given instance,
a setting known as hierarchical multi-label annotation \cite{cartwright_audio_2019, zhang_ontological_2018}.
Given their usefulness in organizing thought, one might expect that leveraging the hierarchy during annotation may yield higher quality and efficiency.
However, there are many design choices to consider: 
    Does interface complexity increase cognitive load (\cite{george_saade_first_2007, oviatt_human-centered_2006})?
    Will false negatives in an upper level of the hierarchy end up amplifying errors into annotations on a lower level~\cite{cartwright_audio_2019}? 
    Will presenting one part of the hierarchy while hiding the rest create a lack of context that leads to misinterpreted label definitions?

In this paper, we study how to incorporate the concept hierarchy into labeling schemes for crowdsource data annotation platforms such as Amazon Mechanical Turk (AMT).
We focus on a difficult annotation task, assigning vaccine concerns (\cite{vax_taxonomy}) to text passages taken from anti-vaccination websites.
The hierarchy of vaccine concerns includes labels such as \textit{``Health risks''} and \textit{``Issues with research''}.
\new{
Small, purpose-built taxonomies are common in the domain of misinformation research \cite{coan_climate_2021, jacobson_taxonomy_2007, salminen_anatomy_nodate, alemanno2018counter}.
}
In the setting of difficult annotation tasks, we show that labeling schemes incorpor-
ating hierarchies can help annotators perform better 
against ground-truth labels.

We investigate two separate design choices when annotating a single passage: 
    (1) how to \textit{format} the hierarchical labels when shown on the interface, and 
    (2) the \textit{pass-logic} that decides how to coordinate multiple workers towards labeling that passage. 
We compare two formats for presenting the hierarchy to annotators (see Figure \ref{fig:format}):
\begin{itemize}
    \item \textit{multi-label}, which simply presents labels as a flat multiple choice list of options
    \item \textit{hierarchical multi-label}, which presents the entire hierarchy directly to the worker who then marks all relevant labels
\end{itemize}
For pass-logic, we look at three options (see Figure \ref{fig:passlogic}):
\begin{itemize}
    \item \textit{single-pass}, where all labels are presented to a single worker, who annotates the passage on their own
    \item \textit{multi-pass}, which combines multiple workers' annotations for a single passage by partitioning the labels into groups (each worker focuses on a small subset of labels at a time).
    \item \textit{hierarchical multi-pass}, in which a preliminary stage of annotation determines if child-labels need to be annotated.
\end{itemize}
We compare all valid combinations of these formats and pass-logic options under a fixed-budget setting, which provides practical insights for research and engineering teams interested in data collection of hierarchical multi-label tasks. For multi-pass logic, we consider both randomly partitioning labels into smaller subsets or utilizing the groupings given to us by the hierarchy. Our results point to a few statistically significant factors:

\begin{enumerate}
    \item Grouping similar concepts together: When partitioning labels using the hierarchy as opposed to a random partition, we see significantly better performance for the groupings informed by the hierarchy (F1 score of $0.50$ grouped vs $0.34$ random)
    \item Relative performance boost on difficult examples: Explicit access to the hierarchy increases workers performance on more difficult questions (as much as a $+0.40$ in F1 as compared to multi-label).
    \item Boosting true positive frequencies: By filtering out irrelevant passages from stage 2 annotation, more of the examples shown to workers are therefore true positives, which we show is associated with better precision without a detriment to recall. The performance boost from this alone moves the F1 score from $0.50$ to $0.57$.
\end{enumerate}

Our results lead us to believe that difficult, high-subjectivity labeling tasks warrant new recommendations separate from crowdsource design guidelines in previous work (\cite{cartwright_audio_2019, humphrey_openmic_2018}).
We recommend considering incorporating hierarchies into the labeling process, and show a few options for how to do so. This is especially true if optimizing for individual worker performance, while choice of labeling scheme plays less of a role if using aggregation methods across several copies of annotations.

\section{Related Works}
\label{related_works}

The reliance of supervised ML algorithms on labeled data has led to a great wealth of knowledge regarding efficient data labeling at large scale. 
Huge datasets have been constructed requiring immense human labeling time across many media.
Among them are image and video datasets generally containing thousands of classes such as
    ImageNet (14M images) \cite{imagenet}
    and OpenImages (9M images) \cite{kuznetsova_open_2020},
    but even with fewer classes, such as CelebFaces labeling 40 facial attributes (200k images) \cite{celeb_faces}.
Also audio datasets, typically with hundreds of classes, for instance
    AudioSet (2M clips) \cite{gemmeke_audio_2017},
    Free Music Archive (100k clips) \cite{defferrard_fma_2017},
    and OpenMIC-2018 (20k clips) \cite{humphrey_openmic_2018}
Lots of work is focused on allowing this scale of data collection while maintaining high quality \cite{deng_scalable_2014, krishna_embracing_2016, chen_efficient_2010, vondrick_efficiently_2013} or protecting crowdsource workers \cite{irani_turkopticon_2013, barbosa_rehumanized_2019}.

Often, this labeling is done on tasks with low ambiguity or subjectivity, and minimal required training -- which makes them suitable for large scale collection.
For example, in ImageNet \cite{imagenet}, labels are the names of well-known objects such as ``ambulance'', ``folding chair'' or ``snail.''
Even in more difficult audio-annotation tasks such as labeling noise categories in a busy city (\cite{cartwright_audio_2019}), the labels (``jackhammer'', ``car horn'') have strong, objective definitions.

Given the clarity on such label definitions, previous studies on user interface design for crowdsource annotation have recommended increasing annotation throughput, \new{or the rate at which labels are collected from the annotation platform} \cite{findlater_differences_2017, morris_subcontracting_2017, barbosa_rehumanized_2019}. 
\new{Throughput can be very quick for some tasks (minutes for hundreds to thousands of annotations), while other tasks may be much slower.}
Prior work found that single-pass methods have up to 9 times higher throughput if annotations are required to be fully labeled (assigning a value for every label) rather than sparse~\cite{cartwright_audio_2019}.
However, work in psychology has long known that there is a tradeoff between speed and accuracy for any information processing task a human performs \cite{wickelgren_speed-accuracy_1977}.
Other HCI work also studies this tradeoff \cite{mackenzie_lag_1993, zhang_text_2019}.
This suggests optimizing for throughput could be harmful to annotation quality, particularly if the task is difficult.

The cognitive load theory \cite{sweller_chapter_2011} suggests that tasks with high cognitive load (the amount of mental effort) can induce errors and mistakes at higher frequency than tasks with lower cognitive load.
Work on user interfaces which require some level of accuracy often tries to minimize unnecessary cognitive load \cite{vondrick_efficiently_2013, george_saade_first_2007, oviatt_human-centered_2006}.
Similarly, work in crowdsourcing recommends to chunk difficult tasks into smaller units of work \cite{kittur_crowdforge_2011}.
Some work has shown that crowdsource platforms have great potential for rapidly collecting measurements in user studies \cite{kittur_userstudies_2008}.
Other work examines how long annotators remain on tasks, and characterizes differences between those that annotate few examples versus those that annotate many \cite{eveleigh_dabblers_2014}.

Recent efforts have also moved towards datasets for high-impact social issues such as:
        misinformation \cite{coan_climate_2021, vax_taxonomy}, which attempts to classify common concerns regarding issues such as climate change or vaccines;
        fact-checking \cite{thorne_fever_2018}, which labels whether claims are verified by trusted sources;
        and claim review \cite{arslan_benchmark_2020, arslan_modeling_nodate}, which determines if claims are worth fact-checking.
Such labels inherently lend themselves to be a more difficult annotation task, given the subjective label definitions and necessary processing to parse written rationales or arguments in text.

In data labeling, it is common to collect multiple copies of annotations and aggregate them using a majority vote \cite{multinli, bowman_snli_2015}.
\new{
Some work studies how to perform aggregation more effectively \cite{whitehill_whose_2009}.
This is said to reduce the impact of low-quality annotations during collection.
Some old work in aggregation methods such as EM uses weightings from estimates of worker skill \cite{dawid_maximum_1979}, while other work incorporates question difficulty 
 through parametric approaches \cite{khetan_achieving_2017} or non-parametric approaches \cite{shah_permutation-based_2021}.
}

However, recent trends in NLP have began questioning aggregation, arguing that subjective labels should not be aggregated if multiple opinions are valid. 
Rather, this line of work (\cite{nie_chaos_2020, zhou_distributed_2021}) suggests predicting the distribution of human opinions, rather than the majority vote.
One implication that follows is that individual annotator performance becomes more important, since one cannot aggregate away labeling error using a simple majority vote.

Labels are not always in the form of lists.
There has been a large amount of work on labeling hierarchical multi-label annotations \new{\cite{cartwright_audio_2019, zhang_ontological_2018, humphrey_openmic_2018, sigurdsson_much_2016}}, where the task is to select any relevant option from labels in a hierarchical structure.
While most work employs a small group of experts to build the concept hierarchies before it gets labeled by workers, some research attempts to build these hierarchies through crowdsourcing methods \cite{chilton_cascade_2013, bragg_crowdsourcing_2013}.

In considering the performance of crowdsource workers, a lot of effort has been spent to introduce gamification of the labeling task \cite{iacovides_games_2013, ohn_evaluation_2019, seong_designing_2020, miyata_gamification_2022, lee_greenify_2013}, but we note that this requires significant overhead efforts to build the games, which may not be feasible when data collection is time-sensitive.

\section{Approach}
\label{approach}

\subsection{Data collection}
\begin{figure}[H]
   \centering
   \includegraphics[width=.4\textwidth]{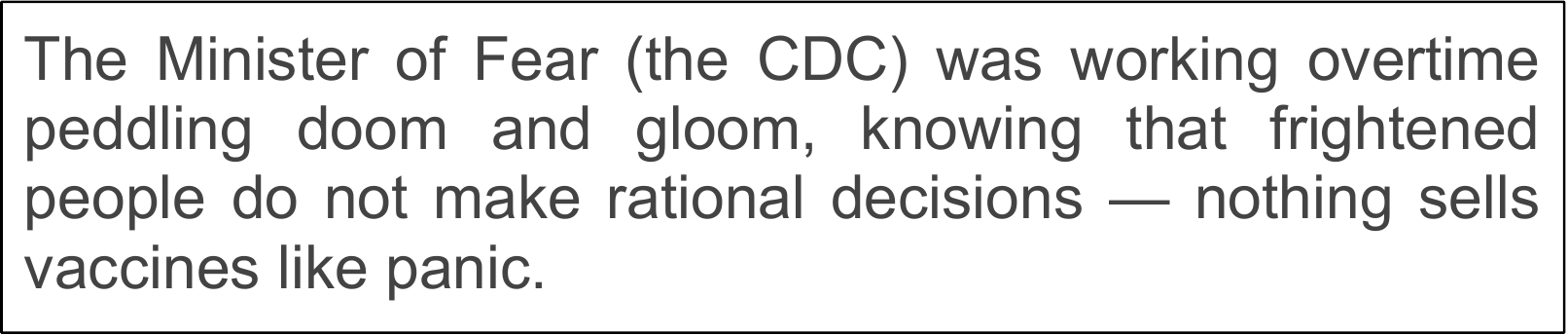}
 \caption{
    Example passage from an anti-vaccination blog.
    Here, the correct labels from the taxonomy include \textit{Untrustworthy actors $\rightarrow$ Profit motives} since the mention of ``selling'' implies money is a corrupting motive, as well as \textit{Lack of benefits $\rightarrow$ Insufficient risk} since ``peddling doom'' implies that the dangers of the disease are being exaggerated.}
\Description{The figure shows an example passage that reads: 'The minister of fear (the CDC) was working overtime peddling doom and gloom, knowing that frightened people do not make rational decisions --- nothing sells vaccines like panic.'}
 \label{fig:hard_example}
\end{figure}

We study interface designs when labeling against a taxonomy of vaccine concerns developed to promote high agreement among crowdsource workers \cite{vax_taxonomy}.
The taxonomy is a hierarchy of labels with $5$ top-level concerns such as \textit{Untrustworthy actors} and \textit{Health risks}, and $19$ child labels such as \textit{Untrustworthy actors $\rightarrow$ Profit motives}. See Appendix \ref{apx:taxonomy} for the full version of the taxonomy.
The annotation task consists of annotating passages
from known anti-vaccination blogs and websites, pre-filtered to ensure the articles are on the topic of vaccination, against multiple labels from both levels in the taxonomy.
Articles are converted into paragraphs using existing markers in the HTML code to closely resemble the paragraphs rendered to readers.
An example is shown in \autoref{fig:hard_example}.

These blog articles are often written with vague mentions of these recurring themes of concerns, and paragraphs are given to annotators without context regarding who wrote it or what paragraph came before.
There is therefore lots of ambiguity in the input text which must be dealt with by annotators.
The authors had some disagreement initially in 45\% of passages, an indication of the level of subjectivity existing in the task.
This is not surprising, given the labeling task primarily revolves around a concept ripe with subjectivity: concerns.
\new{Passages may simply raise different concerns for different readers.}
Unlike the annotation of object in images, for example, there are very few passages where the correct labels are immediately obvious.
That being said, such passages do occur---particularly when there is high overlap between the vocabulary used to define labels and the vocabulary in the passage.

\subsection{Annotator training}
\label{sec:training}
To maximize the chance of high-quality annotations, we look into a few methods to train annotators and ensure quality.
These methods are implemented through the exact same process for all labeling schemes to ensure fairness.
We collect all our annotations on Amazon Mechanical Turk (AMT).
\subsubsection{Definitions}
We provide written definitions for all labels and set up a micro-task as the very first step to have workers interact with the definitions directly.
The very first screen the annotators will see is a list of all the labels they are expected to select from.
Under each label is a written definition.
The task we ask workers to complete is to mark any definition which they feel is unclear.
This hopes to prompt fully reading and internalizing the definitions, as well as collects data for further improving the training process.\footnote{For this paper we do not alter the training process in order to control for this step across all labeling schemes}
(See Appendix \ref{apx:definitions_task} for a screenshot of this step)
\subsubsection{Tutorial}
Next, workers walk through 10 examples, where they annotate passages just like they would in real annotation.
However, for these 10 examples, they are given corrections after each submission.
The corrections show which labels they got wrong and which they got correct.
For incorrectly marked labels, there is a written explanation for why the label should have been applied (or not).
Tutorial explanations are written ahead of time, and appropriate tutorial examples are given according to which labels are presented to the worker.
We ensure that there is always a consistent ratio of different types of examples in each tutorial.
For example, there are always two examples where none of the labels should be selected, one where the passage is clearly anti-vaccination but no specific argument is made (e.g., ``vaccines are bad'') and one where the passage is off-topic.
\subsubsection{Entrance exam}
After finishing the tutorial but before being allowed to annotate real data, we have workers complete a three-question entrance exam. 
To workers, this looks like regular annotation.
Two of these passages are clearly off-topic, and a third passage clearly mentions one of the concerns being labeled.
If workers fail any one of these three questions they are banned from labeling.
\subsubsection{Quality checks}
While the annotations are being collected, we randomly include attention checks (with $5\%$ probability) such as 
    ``Help us catch cheaters. Choose the first option and hit submit to show you are paying attention.''
If workers fail such attention checks, we throw out all the annotations they gave us since the last passed attention check, and ban them from further annotations.

\subsection{Ground-truth labels}
\label{sec:ground_truth}
To evaluate the different labeling schemes, we collect ``expert'' annotations from
three authors of the paper.
\new{The sample size for evaluation spans 4,800 passage-label pairs (200 passages taken from 200 articles).}
First, the three authors annotate the passages separately,
followed by a discussion phase in which they try to come to an agreement
about diverging labels.
We refer to these labels as the \textit{ground truth}, and separate
them into four categories:
    (1) labels which were agreed on immediately during individual, non-communicative annotation; 
    (2) labels which were agreed on after re-annotating them individually without communicating, but asking for a written rationale for the given label;
    (3) labels which were agreed on after collaborative discussion;
    and (4) labels which never reached unanimous agreement, but rather a majority vote was taken.
These categories can be seen as a proxy for difficulty, requiring increasing amounts of nuanced examination of the target passage. 
\new{Further details on the construction of these sets can be found in Appendix \ref{apx:difficulty_details}}, and an analysis of the effect of difficulty in \S\ref{sec:difficulty}.

\subsection{Labeling schemes}
In this section, we discuss the definitions of each interface design through the two formats we consider 
    (\textit{multi-label} and 
    \textit{hierarchical multi-label}) 
    but also a third option which we do not include in experiments due to prohibitive costs (\textit{binary-label}).
We then explore the three pass-logic options 
    (\textit{single-pass}, 
    \textit{multi-pass}, and 
    \textit{hierarchical multi-pass}) 
and show our design approach for combining these options.

\subsubsection{Formats}
\begin{figure}[H]
   \centering
   \includegraphics[width=.45\textwidth]{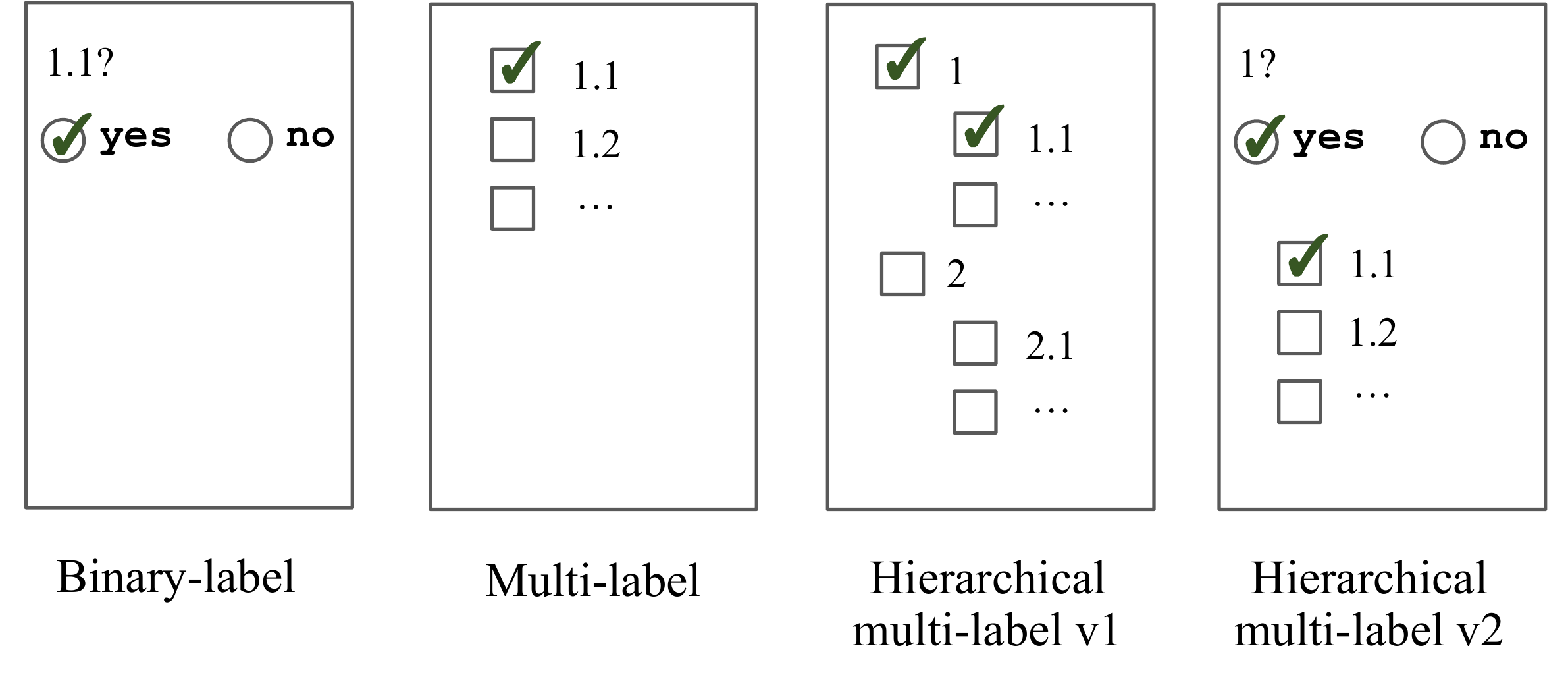}
 \caption{
    User interface designs for presenting labels to the user. 
    In single-pass schemes, using majority vote and the hierarchical multi-label v1 performs the best with a \new{mean} F1 score of 0.70.}
\Description{Four diagrams are shown side by side. In each diagram there are a set of checkboxes or radio buttons indicating how the labels will be presented to the user. Binary label (the leftmost diagram) contains a simple question '1.1?' and below it a radio button reading 'yes' or 'no'. Multi-label contains a simple list of checkboxes labeled '1.1, 1.2, ...'. Hierarchical multi-label v1 contains staggered checkboxes where the leftmost checkboxes read '1, 2' and the boxes immediately under these are tucked under them, reading '1.1, ...' for the parent label '1', and '2.1, ...' for the parent label '2'. Hierarchical multi-label v2 contains both the radio button setup from the leftmost diagram, as well as the checkboxes from multi-label underneath them.}
 \label{fig:format}
\end{figure}

Labels can be presented on an interface in many different formats (see Figure \ref{fig:format}). Here, formats refers to how to organize the set of labels in the user interface.
\begin{itemize}
    \item \textit{Binary-label} format shows the label to annotators using a single yes/no question.
        A worker will focus on a single label across their time annotating, minimizing cognitive load.
            \footnote{This approach has shown useful for high-quality data annotation for images \cite{lin_microsoft_2015}, but has been less successful in video and audio \cite{cartwright_audio_2019, vondrick_efficiently_2013} due to its high cost when there is a temporal element in the annotation.}
    \item \textit{Multi-label}, which simply presents labels as a flat multiple choice list of options.
        The workers can select any/all/none of the labels.
        Depending on the pass-logic used, this list may be longer or shorter, but will generally only contain labels from the same level of the hierarchy.
    \item \textit{Hierarchical multi-label}, which presents a hierarchy directly such that choices in the top-level of the hierarchy prompt further choices in the next level.        This option can be accomplished in two ways. 
        In one version (v1), the hierarchy is given as checkboxes with child-level checkboxes that become enabled only if the parent category is selected.
        In the other (v2), the hierarchy is asked in a two-stage question. First there is a binary choice regarding the parent category. If the answer is yes, then a flat list of checkboxes for the child labels is presented to the worker.
\end{itemize}

\subsubsection{Pass logic}
Pass logic determines how many workers are brought together to work on the annotation of a single passage, and how to coordinate their efforts.
Some decisions regarding pass logic will inform the look of the interface shown to users, while others will only affect which passages are shown to any given worker.
We examine three options for pass logic (see Figure \ref{fig:passlogic}):
\begin{itemize}
    \item \textit{Single-pass}, where one worker is asked to annotate the passage entirely on their own, and therefore must be presented all the labels at once.
        This option can be combined with both the \textit{multi-label} and \textit{hierarchical multi-label} formatting options. 
        However, it is incompatible with \textit{binary-label} since you cannot present multiple binary questions at once (that would be \textit{multi-label}).
    \item \textit{Multi-pass}, which combines the annotations of multiple workers for a single passage by partitioning the labels into groups (letting each worker focus on a small subset of labels at a time).
        This option is compatible with all format versions.
        To accomplish this with the \textit{hierarchical multi-label} format, we simply partition the hierarchy into sub-trees using the top-level labels.
    \item \textit{Hierarchical multi-pass}, in which there are different stages of annotation which determine whether child labels in the hierarchy need to be annotated.
        First, some worker is asked to annotate the passage with level-1 annotations.
        Based on their annotations, we create new tasks for any label the worker marked as positive. 
        These new tasks are released in a second stage to annotate the child labels in level 2, and need not be labeled by the same worker as the level-1 labels.
        This option can be employed both in a \textit{binary-label} setup, as well as in \textit{multi-label}, but is incompatible with \textit{hierarchical multi-label} formatting since it forces annotation to occur on distinct levels at a time.
\end{itemize}

\subsubsection{Combinations}
\label{sec:labeling_schemes}
When combining the format options with pass-logic options, we get the following possible labeling schemes:

\begin{itemize}
    \item \textbf{Single-pass multi-label} \new{(single-pass multi)}
        --- A single worker annotates all level-2 labels at the same time, given in a flat list.
    \item \textbf{Single-pass hierarchical multi-label} \new{(single-pass hrchl)}
        --- A single worker annotates all labels (level-1 and level-2) at the same time. 
        They are shown the hierarchy in its entirety using hierarchical multi-label v1 formatting (see Figure \ref{fig:format}).
    \item \textbf{Multi-pass binary-label} \new{(multi-pass binary)}\footnote{Binary label schemes are not included in experiments due to their high cost.}
        --- The labels are given one by one to multiple workers, who annotate the passage in parallel for that single label. 
        Annotations from all workers are then combined.
    \item \textbf{Multi-pass multi-label}  \new{(multi-pass multi)}
        --- Level-2 labels are partitioned into smaller groups, and these label-groups are then given to multiple workers who annotate the passage in parallel.
        Annotations from all workers are then combined.
    \item \textbf{Multi-pass hierarchical multi-label} \new{(multi-pass hrchl)}
        --- The labels are partitioned according to level-1 labels.
        One worker will be given label 1 and its children 1.1, 1.2, ..., while another worker will be given label 2 and its children 2.1, 2.2, ... 
        This pattern continues for all level-1 labels in the hierarchy.
        They are shown their section of the hierarchy using hierarchical multi-label v2 formatting (see Figure \ref{fig:format}).
    \item \textbf{Hierarchical multi-pass binary-label} \new{(hrchl-pass binary)}
        --- Level-1 labels are first given one by one to multiple workers, who annotate the passage in parallel for their single label. 
        A second stage then looks at these annotations and determines which child labels need to be labeled (if a worker indicates a positive label for label 2, then we must annotate 2.1, 2.2, ..., else we can skip them).
        The second stage then gives the child-labels one by one to multiple workers in parallel, just like the first stage.
    \item \textbf{Hierarchical multi-pass multi-label}  \new{(hrchl-pass multi)}
        --- Level-1 labels are first given as a single group to one worker.
        A second stage then looks at this worker's annotations and determines which child labels need to be labeled (according to the same logic as in the binary case).
        The second stage then gives the child-labels in groupings according to their parent category (so a single worker will be given 2.1, 2.2, ... at once), just like in the first stage.
\end{itemize}

When partitioning the level-2 labels for \textit{multi-pass multi}, we examine two possible choices: 
partitioning them using the groupings that already exist in the hierarchy, or partitioning them randomly into 5 groups (such that the number of groups is consistent with the other choice). 
\new{We refer to these as \textit{multi-pass \textbf{grouped} multi} and \textit{multi-pass \textbf{random} multi}, respectively.}

\subsection{\new{Controls}}
\new{
Beyond forcing an annotator training, we explore several additional controls. This section outlines the controls we took, and which factors we look at through post-hoc analysis.
\S\ref{sec:cost} looks at how we control cost, which is key to our experimental design.}

\new{
Given the task difficulty, we limit access to the task to workers who (1) reside in the United States, (2) have completed at least 2,000 HITs, and (3) have a HIT approval percentage of above 99\%, and (4) have a ``Masters'' qualification indicating they are workers that produce high quality annotations.
These controls are facilitated by standard AMT tools, while most of the rest of the controls are implemented through our custom annotation platform.
}

\new{
We use a between-subjects design, meaning that we do not allow any worker to submit annotations for more than one labeling scheme. This avoids producing workers which are trained twice on the task. 
Further, the workers are not aware that there are multiple conditions.
When publishing jobs on AMT we start with HITs that will send the workers to the first labeling scheme. Once we have collected enough annotations for this scheme, the current workers get blocked from beginning any new hits. The next labeling scheme then gets linked from the posted HITs, and new workers (which did not interact at all with the first labeling scheme) may begin annotation.
This ensures workers are not aware of multiple schemes, even if they have seen the HIT advertised in their list of tasks previously. The description of the task is the same, except for the reward which fluctuates slightly to maintain a consistent budget (more details in \S\ref{sec:cost}), and we never inform the workers that there are multiple schemes.
Workers are at most allowed to submit annotations for 200 unique passages.
}

\new{
We do not control exactly when these HITs are submitted to the AMT marketplace. Simply, we launch the next HITs shortly (1-2 hours) after gathering enough annotations for the last labeling scheme. When the last labeling scheme finishes collection during the night or late in the evening, we wait until the morning to launch the next scheme.
One could argue that the populations of workers that click on tasks might vary meaningfully across the day. 
However, most annotations were collected during day-time in the United States, and we only allow workers from the United States.
We also include this factor in our multiple regression analysis in \S\ref{regression}, showing it does not significantly contribute to mean F1 score.
}
\new{
Since we allow workers to complete any number of annotations they want (up to 200 unique passages), we cannot control how many workers are assigned to each condition.
Instead, we allow annotation by new workers up until we have 3 copies of each of the 4,800 passage-label pairs.
See Table \ref{fig:nr_labelers} for more details.
}

\begin{table}[H]
 \begin{tabular}{@{}c@{}} 
  \includegraphics[width=.48\textwidth]{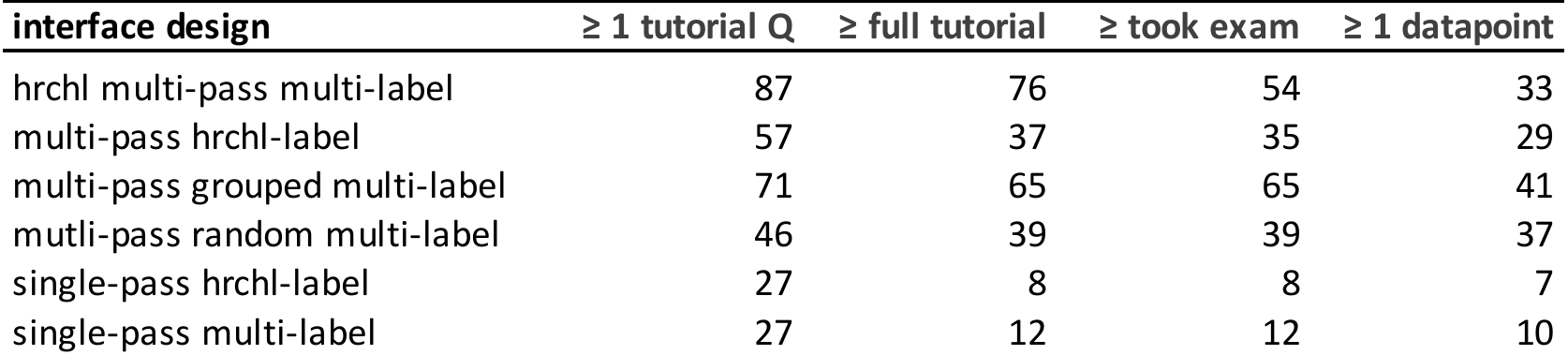}
 \end{tabular}
 \caption{
    \new{
    Number of workers which began each stage of the data collection pipeline, broken down by labeling scheme.
    Since multi-pass schemes required more annotations, workers had more time to reserve HITs and submit annotations before it had been completed. This has some affect on our confidence intervals for single-pass schemes.
    }
    }
 \label{fig:nr_labelers}
\end{table}

\subsubsection{Cost}
\label{sec:cost}
To control for cost, we approach the task from the perspective of a research team that wants to collect data to train an ML model.

Note that annotating text passages is very different from images.
Whereas images can be cognitively processed near-instantaneously by a crowdsource worker, reading passages of text is more similar to the annotation of video or audio-clips.
Performing a full read-through of the passage takes time, forcing a delay before selecting labels and thereby adding a significant temporal dimension to the annotation task.
Therefore, we cannot compare 
    the annotation of 10 passages with a single binary label
    to the annotation of 1 passage with 10 labels.
In one case, the annotator has to spend 10 times more time reading than the other. 
This influences how to fairly pay workers.
Instead, we must set a minimum reward threshold \textit{per passage read-through} for workers, and consider the cost of data collection as variable.
\new{
For a toy example of why cost would vary across labeling schemes, see Appendix \ref{apx:cost}.
}

\new{
However, comparing labeling schemes without holding the total budget constant will not provide much value for ML researchers deciding which scheme to use. 
}
Research teams are heavily motivated by budgets, so how do they get the highest quality annotation for their money?%
\footnote{\new{We should note that one possibility is to spend the same amount of money but vary the amount of data collected. This adds complexity to this question beyond the scope of this paper, since it would be necessary to build and evaluate ML model performances in order to measure the tradeoff of data \textit{quantity}.}}
Answering this question is the focus of our experiments (\S\ref{analysis}).
In particular, we set the reward per passage read-through for each labeling scheme to fully utilize the budget
(ensuring that it was above a minimum of $\$0.10$, which ensures that we are paying workers more than the United States minimum wage).
This means in some labeling schemes the workers will get more rewards per passage than others, although these workers also have to consider more labels at the same time.

\new{
On AMT, workers are paid per HIT (one ``unit'' of work assigned to a worker) they complete.
For each HIT, workers in our experiment will complete a small batch of passages (10-24 passages).
Batching passages like this ensures the reward per HIT is not too small to attract workers.
This also allows us to change the ratio of reward-to-number-of-passages, thereby controlling for the listed reward payment per HIT that workers see on the platform.
We observe that changing this ratio (without changing the actual payment per work completed) causes noticeable differences in annotation throughput.
This indicates a potentially large inefficiency in the AMT marketplace.
For all labeling schemes, we launch the tasks with a ratio of reward-to-number-of-passages such that the reward is just above a dollar (as close to \$1.01 as we can get while keeping the budget fixed).
For all schemes we still keep the total budget spent constant for collecting the 14,400 labels needed (3 copies of 4,800 passage-label pairs).
}

\subsubsection{Payment broken down by each labeling scheme}

From the process described in \ref{sec:cost}, we then end up with the following rates of pay for each condition:
\new{
The listed reward for \textit{hrchl-pass multi} was \$1.01 for 10 passages, all \textit{multi-pass} options were \$1.03 for 24 passages, and single-pass options were \$2.16 for 10 passages (after first having tried \$1.08 for 5 passages and finding throughput was too slow).
}
\new{
The reward we give workers amounts to approximately \$7-10 per hour (USD) as self-reported through TurkOpticon}%
    \footnote{The TurkOpticon page for our requester account shows 5/5 rating in Fairness and 5/5 rating Fast payments. This requester account was created solely for these experiments.}\cite{irani_turkopticon_2013}.
\new{
We do not have access to more granular hourly-rate estimates due to limitations with monitoring when workers are inactive (taking a break) versus when they are taking longer than usual time to read a question.
However, we include analysis regarding distributions of time spent labeling each passage across the various schemes in Appendix \ref{apx:time_distributions}.
}

\section{Analysis}
\label{analysis}
We collected three copies of annotations for each passage, for each of the 6 labeling schemes (\S\ref{sec:labeling_schemes}) through AMT. 
In this section we 
    compare the labeling schemes against each other on performance and
    examine the reasons for why performance varied across labeling schemes.


\subsection{Performance Comparison}
\label{sec:performance}
We evaluate the performance of workers against the ground-truth labels (\S\ref{sec:ground_truth}).
Majority labels are often computed to mitigate labeling error \cite{whitehill_whose_2009}, but recent work has also shown the utility of high-quality individual annotations in order to estimate the distributions of human opinion \cite{zhou_distributed_2021}.
The latter is particularly relevant in our setting where workers
are labeling often subjective concerns:
being able to measure the degrees of concern across individuals is relevant towards reducing vaccine hesitancy.
We compute the precision, recall and F1 score for each label of the vaccine concerns taxonomy, and report an unweighted mean across the labels. 

\new{
We employ a macro-level average of F1, which is computed by first finding the F1 score on every taxonomy label, and then averaging across all these labels.
Note that in any analysis where we give an individual F1 score for each worker, the macro-averaging process happens in parallel for each worker. That is, the worker would be evaluated separately for each taxonomy label, and then an average performance is computed for that worker.
However, for most of our analyses, we look at a single F1 score across all workers. 
In this case, we first pool all the annotations and treat them as if a single worker had submitted them. 
We then follow the macro-averaging process across the taxonomy labels.
For further details on the metrics we use, see Appendix \ref{apx:metrics}.
}

\new{
We generate a choice/random baseline. For each passage, we draw 3 samples for each label from the binomial distribution with the probability p being determined by the gold-labeled data. We employ the same scheme to ensure consistency as described in Appendix \ref{label_consistency}.
Note that since F1 is computed using a macro average, and since there are ``nans'' in the data when positive labels are not generated, the mean F1 will not necessarily lie between the mean precision and mean recall.
}

\begin{table}[H]
\begin{tabular}{@{}c@{}}
   \includegraphics[width=.4\textwidth]{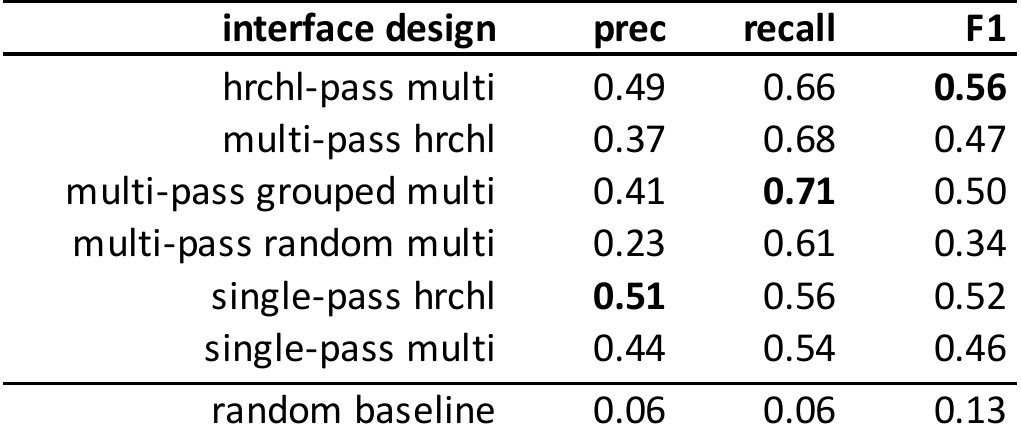}
\end{tabular}
 \caption{Mean performance of crowdsource workers against ground truth labels.
    Hrchl-pass multi-label performs best on mean F1. \new{We include full breakdowns of these F1 scores (and other performance metrics) by label in Appendix \ref{apx:perf_by_label}, as well as confidence intervals for all metrics in Appendix \ref{apx:conf_intervals}}
    }.
 \label{fig:performance}
\end{table}

Workers annotating with \textit{single-pass hrchl} had the highest precision of $0.51$, while \textit{multi-pass grouped multi} had the highest recall at $0.71$.
\textit{Hrchl-pass multi} balanced these the best, with an F1 score of $0.56$. 
Generally, the data indicates that single-pass options lead to higher precision, while multi-pass and hierarchical multi-pass options perform better on recall.

One possible explanation could be that when workers focus on a smaller set of labels, they have a lower chance of forgetting about them while reading the passage.
The tradeoff would be that as workers see a longer list of labels, they have to be more certain the passage is speaking about a label to think of it and select it.
It could also be possible that workers ``want'' to select \textit{something} on each passage.
When the options are few they tend to over-annotate, and when the options are many they find the obvious ones more easily, producing fewer false positives.
There is some evidence for this explanation.
The mean number of selections per passage in the single-pass schemes was $0.9$, while the mean number of selections in multi-pass options was $1.5$, indicating that partitioning the labels into smaller categories may cause workers to annotate more positives than if they are given all together.

\subsection{Multiple Regression Analysis}
\label{regression}
\new{
We investigate the effects of various factors on worker F1 scores. 
In this section, we fit a multiple regression model to the F1 scores of each worker.
See Table~\ref{fig:nr_labelers} for how many workers completed annotations in each labeling scheme.
We consider several factors beyond the labeling scheme, including ones that were not controlled for in our experimental design (such as the time each labeling scheme was distributed on AMT) as well as factors which arise due to each worker's ``luck'': the percentage of passages they were given which were relatively easy, and how often they were shown a passage which \textit{should} be labeled with some positive label.
}

\new{
The \textit{labeling scheme} factor is a categorical variable encoding the 6 labeling schemes considered in this paper. \textit{Multi-pass random multi} is set as the baseline for this analysis.
\textit{time started} is a variable encoding when during the day a given worker began annotating passages. It is given in seconds past midnight.
\textit{percentage easy/medium/hard/no agreement} factors encode the percentage of easy / medium / hard / or no agreement (referring to the 4 proxy levels for difficulty) labels which the given worker was presented with. Some workers, by luck, get easier or harder passage-label pairs shown to them, and here we hope to see what the effect of this is. Details on how we choose these 4 difficulty levels are given in \S\ref{sec:difficulty} and Appendix \ref{apx:difficulty_details}.
\textit{true pos freq} is a variable encoding what percentage of labels shown to a given worker \textit{should} be labeled positively.
}

\new{
F1 scores were significantly improved by three labeling schemes above the baseline: multi-pass grouped multi-label (estimate = 0.13, \textit{p}-value < 0.001), single-pass hrchl-label (estimate = 0.08, \textit{p}-value < 0.05), and single-pass multi-label (estimate = 0.07, \textit{p}-value < 0.1).
}

\new{
For factors beyond the labeling scheme, we see that 
    the time of day each worker began the task did not have a statistically significant effect on the data (\textit{p}-value = 0.66), whereas both 
    the rate of true positives that workers encounter during annotation (estimate = 0.54, \textit{p}-value < 0.001) and 
    the percentage of easy passages they encounter (estimate = 0.39, \textit{p}-value < 0.05) do have a statistical significance.
This is especially of interest to us since these factors can be indirectly manipulated through the labeling scheme. We analyse these factors in further detail in \S\ref{sec:difficulty} and \S\ref{sec:tp_freq}.
}

\begin{table}[H]
 \begin{tabular}{@{}c@{}}
   \includegraphics[width=0.48\textwidth]{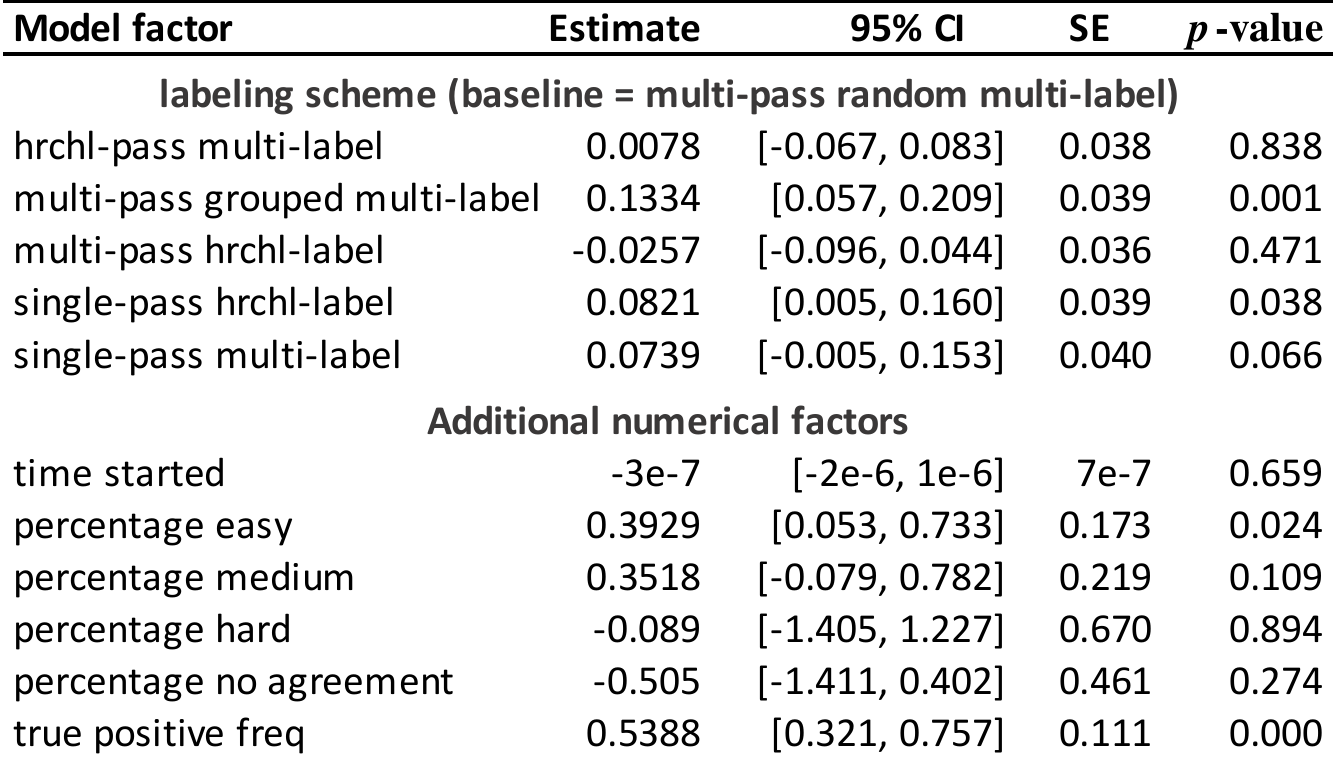}
 \end{tabular}
 \caption{\new{Multiple regression analysis for factors influencing worker F1 score on a per-label basis. easy / medium / hard / no agreement refer to the 4 categories described in \S\ref{sec:ground_truth}. Note that a coefficient of 0.07 can indicate the difference between (for example) 0.50 and 0.57 F1 scores, since labeling scheme factors are coded as binary against the baseline scheme.}
 }
 \label{fig:regression}
\end{table}

\subsection{Contributing factors toward performance differences}
In this section we perform deeper analysis on potential reasons for performance differences across labeling schemes.

\subsubsection{Grouping labels}
Overall, integrating the hierarchy into the labeling scheme seems to help with performance.
One direct comparison we can make is between the two versions of \textit{multi-pass multi-label} schemes. 
In one, \textit{multi-pass random multi}, the level-2 labels are partitioned randomly and given to separate workers.
In \textit{multi-pass grouped multi}, we use the groupings that already exist due to the hierarchy.
Comparing performance between these schemes helps us examine whether presenting conceptually similar categories together can boost performance.

In every single measurement (accuracy, precision, recall, F1) and in every single vote setting (sensitive, majority, unanimous), the grouped scheme outperforms random partitions.
\new{On individual workers' mean performance, \textit{multi-pass grouped multi} scores $0.50$ with a 95\% confidence interval of $[0.45, 0.55]$, while \textit{mutli-pass random multi} only scores $0.34$ ($[0.29, 0.43]$).}
It seems important when partitioning the labels to group related labels together.
It is unclear exactly why this is, but one possibility is that having the context of similar labels increases worker's understanding of the nuance between different cases.
If they are shown a passage with a text which has criticism of research, it may be useful to be labeling both \textit{``Issues with vaccine research $\rightarrow$ poor quality''} alongside \textit{``Issues with vaccine research $\rightarrow$ lacking quantity''} rather than just one (without knowledge about the other).

\subsubsection{Examining difficulty}
\label{sec:difficulty}
\begin{figure}[H]
   \centering
   \includegraphics[width=.4\textwidth]{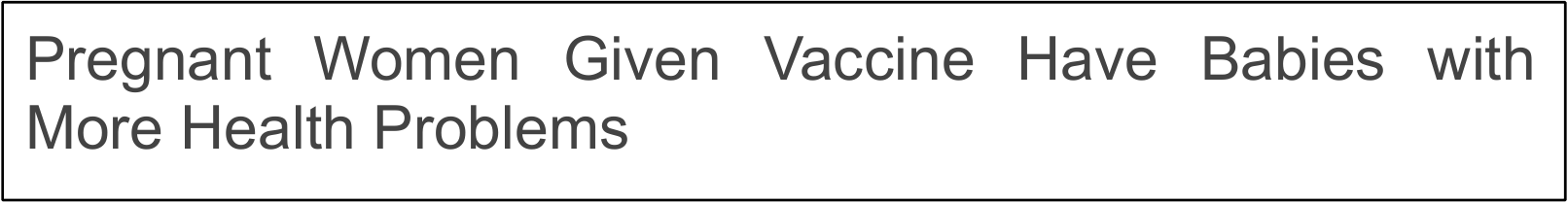}
 \caption{An easy-to-label example passage from an anti-vaccination blog.}
\Description{A table is shown with column headers reading 'interface design', 'greater than or equal to 1 tutorial Q', 'greater than or equal to full tutorial', 'greater than or equal to took exam', and 'greater than or equal to 1 datapoint'. The table shows values for all 6 labeling schemes.}
 \label{fig:easy_example}
\end{figure}

Even though our task \textit{generally} contains more ambiguity and is higher in cognitive load than other crowdsourced annotation tasks, there are of course easy cases to label.
For example, the passage below (Figure \ref{fig:easy_example}) should very clearly be labeled with \textit{``Health risks''}.

We utilize the ground-truth label categories discussed in \S\ref{sec:ground_truth}, and examine the difference in performance as we vary difficulty.
Importantly, we do not simply assign a difficulty measure to each passage, but rather to each passage-label pair.
That means that we are able to mark that it is easy to annotate \textit{``Health risks''} for the passage in Figure \ref{fig:easy_example}, but we can also mark that it is difficult to annotate the label \textit{``High risk individuals''} if that was a label the authors did not immediately agree on.
This analysis is done post-hoc.
The passages are given at random ordering to workers, so workers will in expectation see the same proportion of difficult passages.

\begin{figure}[H]
   \centering
   \includegraphics[width=.45\textwidth]{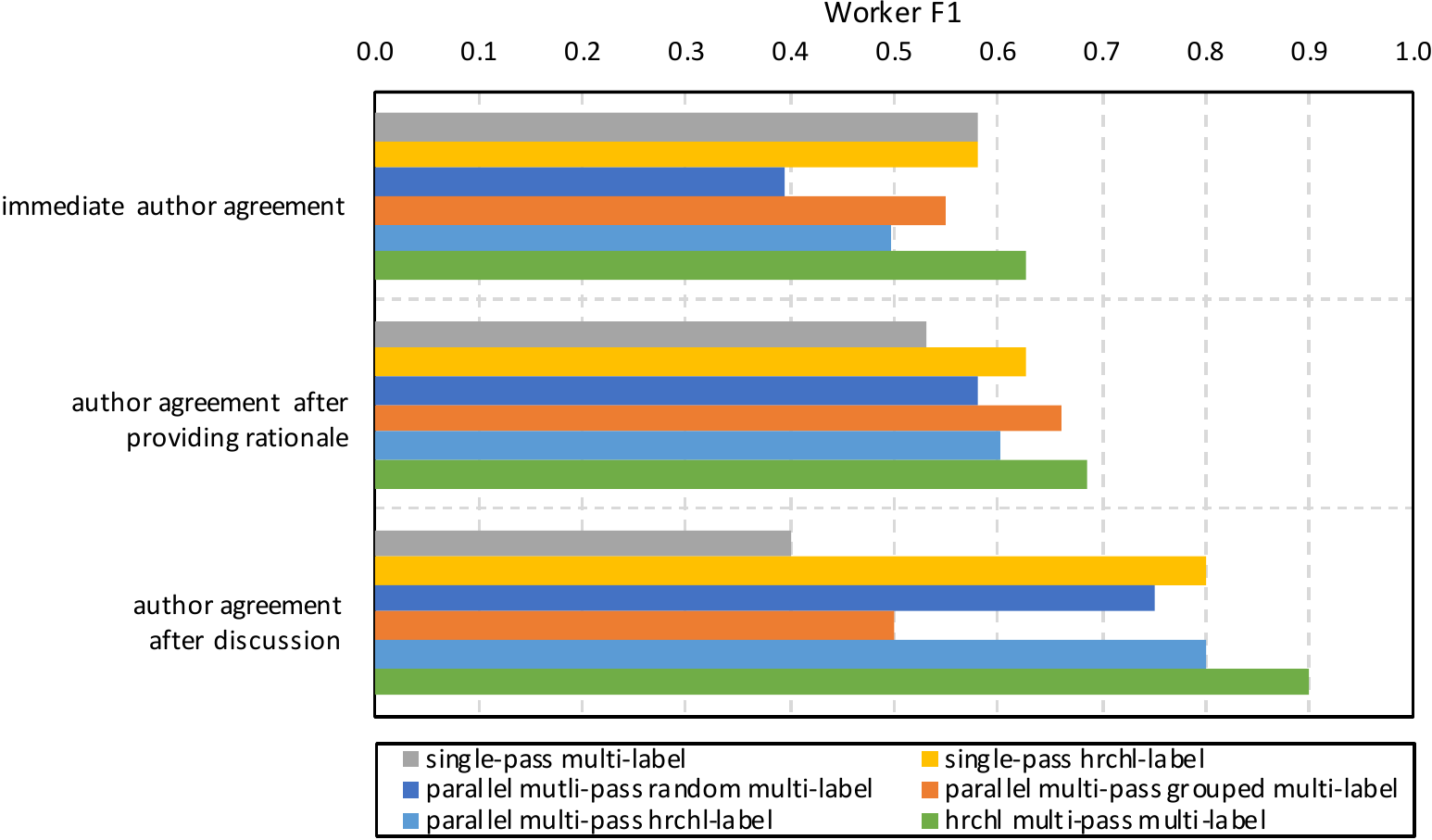}
 \caption{
 \new{
    Worker mean F1 score versus increasingly difficult passage-label pairs. 
    Note that some platforms perform better on more difficult passages, largely driven by their performance in precision.
    Note that the fourth category is excluded from this analysis due to lack of enough data, and semantic issues with how to compare performance against passages where both binary labels are technically valid.
    }}
\Description{A table is shown with values for precision, recall, and F1 score. These metrics are given for each of the 6 labeling schemes, and a random baseline is shown at the bottom. Hrchl-pass multi has bold font at the f1 score indicating it is the highest value in that column: 0.56.}
 \label{fig:difficulty_f1}
\end{figure}

\new{
Appendix \ref{apx:difficulty_details} shows similar plots for accuracy, recall, and precision.
}

Focusing in on two comparable labeling schemes, the two single-pass versions, we see that performance on the labels diverges as difficulty increases.
Performance on the easiest category (immediate agreement among authors) is almost identical (F1 score of $0.581$ for \textit{single-pass multi} and $0.582$ for \textit{single-pass hrchl}), while the difference is already $+0.400$ in the favor of \textit{single-pass hrchl} as we reach the most difficult category where there is still author consensus.
This generally supports the explanation that explicitly providing the hierarchy helps workers reason about difficult labels.
It is unclear, however, exactly why the performance \textit{increases} as difficulty increases for the \textit{single-pass hrchl} scheme. 
It is possible that the tradeoff between the helpful structure and the harmful interface complexity interact such that this labeling scheme performs worse on easier passages.
Alternatively, it may be an effect of correcting workers' priors for assigning a positive label.

\subsubsection{True positive frequency}
\label{sec:tp_freq}
Beyond the interface design format shown to workers, and the pass-logic used to combine annotations, there may be other factors that impact their performance.
Does a worker who sees lots of positive examples perform differently from a worker who rarely sees any positives?

\new{
The results of the multiple linear regression indicates that there is a significant increase in F1 score due to higher true positive frequencies shown to workers.
Knowing this, we may want to design annotation platforms which ``filter out'' negative examples, so that more workers have higher true positive frequencies during annotation. See Appendix \ref{apx:tp_freq_all} for a plot of the relationship between true positive frequency and F1 score among all the workers.
Intuitively, one reason higher true positive frequencies may cause better performance could be that workers \textit{expect} to have to assign positive labels to some proportion of passages, which would cause them to over-assign positives.
}

We examine the performance differences between labels collected in \textit{multi-pass grouped multi} and \textit{hrchl-pass multi}.
If we ignore the level-1 annotations collected in \textit{hrchl-pass multi}, then the interface shown to workers in these two cases is identical.
The only difference is which passages actually get shown.
For \textit{multi-pass grouped multi}, we show all available passages to the workers. There is no pre-filtering on relevant passages done.
For \textit{hrchl-pass multi}, we only show passages that already have a positive annotation of the parent label, meaning there is a high chance of more labels being relevant.
In fact, the frequencies of true positives shown to workers jumps from $3\%$ to $13\%$ on average (a more than 4-fold increase) just from this pre-filtering.
Below, we compare the label performance on the passages that were annotated in both schemes.
\begin{figure}[H]
   \centering
   \includegraphics[width=.45\textwidth]{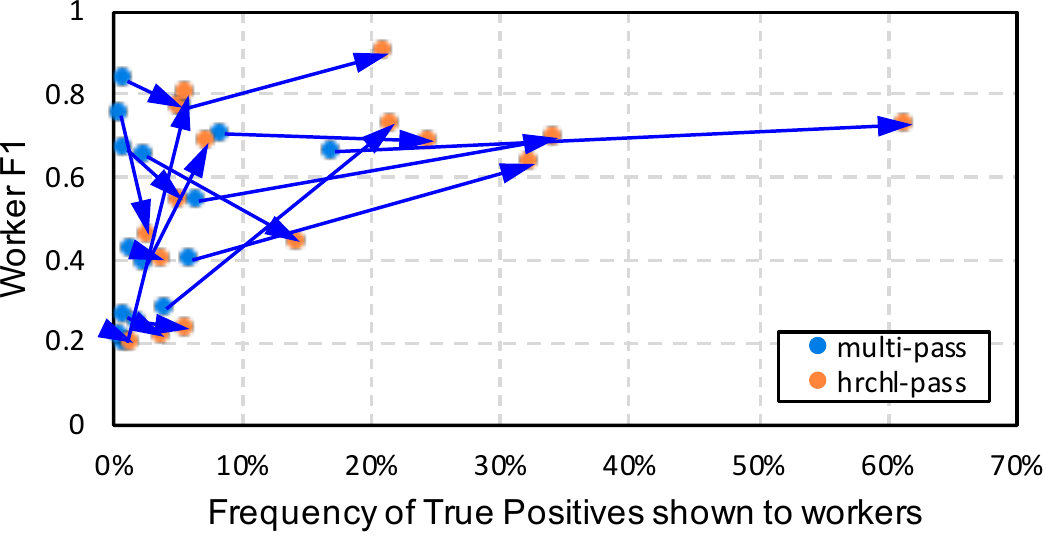}
 \caption{
    \new{
    Worker F1 score as the frequency of seen true positives (TPs) varies. 
    Labeling schemes determine which paragraphs get shown:
        all available (blue) or 
        pre-filtered by the parent label (orange).
    Arrows link the same labels from one scheme to the other.
        }}
\Description{A two-part table is shown with column headers 'model factor', 'estimate', '95\% CI', 'SE', and 'p-value'. The first part of the table (top) has a subheading that reads 'labeling\_scheme (baseline=multi-pass random multi-label)', and the second part of the table (bottom) has a subheading that reads 'additional numerical factors'. The first part includes 5 of the 6 labeling schemes, while the bottom includes new factors such as 'time\_started', 'percentage\_easy', and 'true\_positive\_freq'. Some values in the table are denoted '*' which represents a p-value below 0.001.}
 \label{fig:tp_freq}
\end{figure}
Note that overall we see a statistically significant positive correlation (Table \ref{fig:regression}), when we aggregate across all workers and examine changes on a per-label basis this trend is more nuanced.
Overall, for passages directly annotated by workers in both schemes, \textit{hrchl-pass multi} achieves a mean F1 score of $0.57$ on level 2 labels whereas \textit{multi-pass grouped multi} only scores $0.50$.
This is driven mainly by an improvement on precision ($+6.7\%$) rather than recall, which stays fairly unaffected ($+0.01$).
It therefore seems that better balancing class priors for the workers can help with their performance on the task.
This may warrant recommendations of a pre-filtering step to remove obvious true negatives.
Ultimately, hierarchical multi-pass schemes acts as a form of pre-filtering, and seems to have a positive influence on worker performance.

\subsection{Voting schemes}
\new{
If one's primary goal is not to measure the distribution of judgments about a label, but rather to get a single binary answer for each passage, then employing a vote may still be beneficial.
That is not the primary motivation of this work, but  in order to give some guidance to the implications of our results for aggregation methods, we examine simple, threshold-based voting schemes.
}

We look at three possible vote setting to aggregate the three copies of annotations collected on each passage. 
In sensitive vote, only 1 positive vote (of 3) is required to mark a label as positive. 
In majority, 2 of 3 is required, 
and in unanimous all 3 must be positive to mark it positive.
\begin{figure}[H]
   \centering
   \includegraphics[width=.4\textwidth]{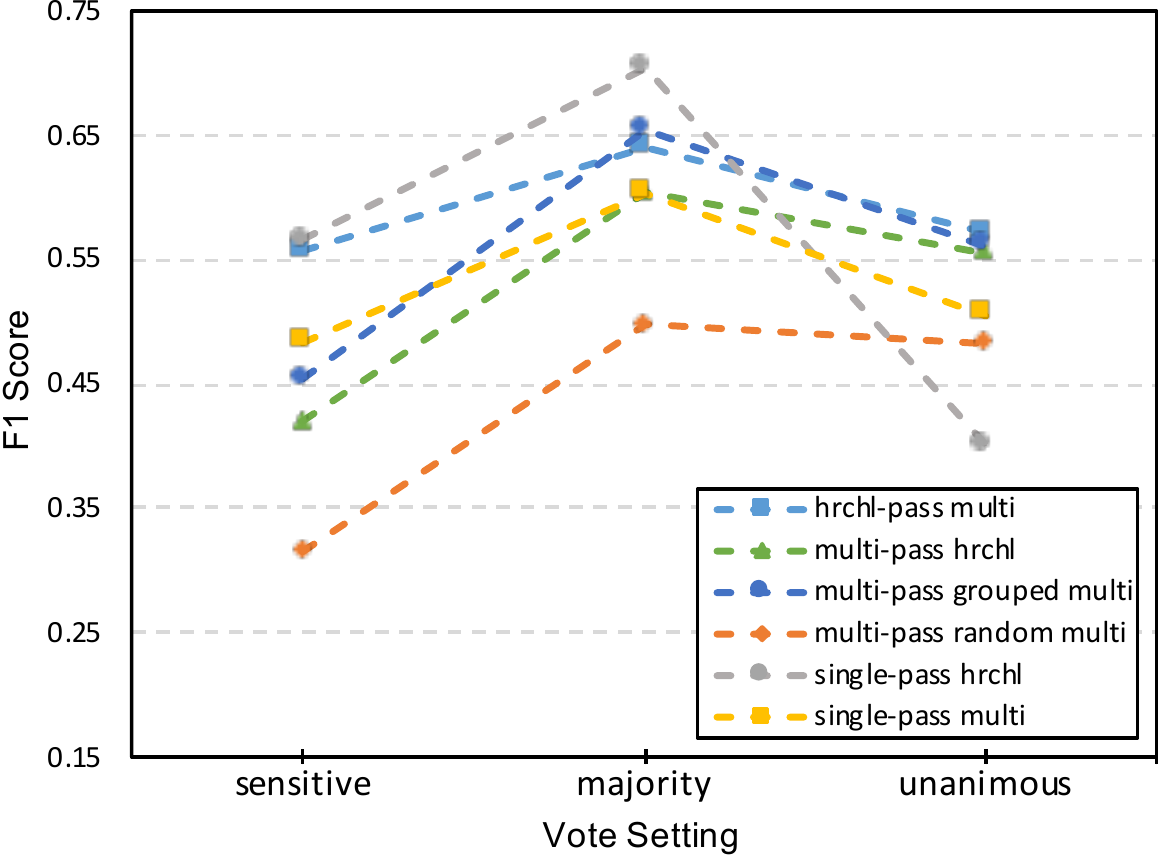}
   \caption{
    F1 score of each vote setting. 
    Consecutive vote settings become increasingly conservative on positive labels, requiring 1, 2, or all 3 votes (respectively) to mark it positive.
    All labeling schemes maximize F1 score using majority score, indicating it is the safest choice for vote setting.
    }
\Description{The figure shows an example passage that reads: 'Pregnant women given vaccine have babies with more health problems'}
 \label{fig:vote_setting_f1}
\end{figure}

Majority vote balances precision and recall the best across all labeling schemes (see Figure \ref{fig:vote_setting_f1}).
Note that as mentioned in \S\ref{related_works}, strong individual annotator performance may still be necessary for some tasks where aggregation is not possible.
Majority vote also outperforms individual worker's performance in all labeling schemes, and the differences in performance between labeling schemes become less pronounced as you aggregate using voting.
The highest F1 score was achieved by \textit{single-pass hrchl} at $0.70$ using majority vote.
Sensitive vote scores the best on recall, with \textit{multi-pass grouped multi} achieving $0.92$, 
and unanimous vote settings perform the best on precision: \textit{single-pass multi} scores $0.93$.
Increasing the vote threshold results in more conservative labeling, which is why we see an increase in precision.

\section{Discussion}
\label{discussion}
There is a large body of work which has studied crowdsource annotations. 
How to make it scalable, how to keep quality high, and how to increase throughput.
There is also a deep wealth of knowledge and best practices regarding user interface design.
Much of this work, however, has been done for labeling tasks that are low-subjectivity, have clearly defined label definitions, and low cognitive load.

We studied six candidate labeling schemes in annotation of a taxonomy of vaccine concerns.
Our motivation was to study whether the hierarchy itself could aid workers as they perform the annotation task, \new{and how to design the annotation task for high-quality labels under a fixed budget}.
We found that integrating the hierarchy into the labeling scheme helps with improving annotation quality, whether explicitly in the interface or through logical passes made behind the scenes.
Our analysis showed that a \textit{hierarchical multi-pass multi-label} scheme performs best when considering individual worker performance.
We believe individual worker performance to carry more importance when the tasks are inherently subjective, since a growing body of work is interested in predicting label distributions.
Much like in \cite{cartwright_audio_2019} \new{and \cite{sigurdsson_much_2016}}, we find that workers assign more labels per passage on average when they are in multi-pass schemes versus single-pass ones.

However, if the priority is to collect high confidence labels rather than distributions of human opinion, we found that employing the \textit{single-pass hierarchical multi-label} along with a majority vote achieves highest performance. 
Unlike \cite{cartwright_audio_2019} and \cite{humphrey_openmic_2018}, we don't see a drop-off in performance when using single-pass labeling schemes.
Although we don't conduct a qualitative study, we did not receive notably different amounts of complaints from workers in any of the labeling schemes.
Largely, complaints were not about the interface designs at all, but rather about being allowed to annotate more data after workers finished the batch or were banned for failing attention checks.
When using majority vote, the choice of labeling scheme matters less than it does for individual worker performance.
The ease of setup with single-pass options should not be undervalued either.
Such labeling schemes are already supported natively in AMT's requester user interface, making it a strong option for smaller projects with a necessary quick turnaround.

Overall, we find that introducing the hierarchy helps almost universally across our experiments.
In comparisons between partitioning the labels into groups randomly versus using the hierarchies structure, we find that using the hierarchy dominates across all performance metrics.
Exposing the hierarchy explicitly helped performance on single-pass schemes by increasing worker performance on particularly difficult passages.
We used a taxonomy specifically designed to achieve high agreement among crowdsource workers.
While some previous work has indicated that integrating hierarchies into the data labeling process may harm performance (\cite{vondrick_efficiently_2013}), we find that it boosts it.
The contexts here are different: our task is higher difficulty and therefore the hierarchy may aid in completing the task, but using a taxonomy specifically designed for high agreement among crowdsource annotations may also indicate that for these methods to work you may need a well-designed hierarchy.

\subsection{Limitations}
\textit{Use of Amazon Mechanical Turk}.
We conduct our experiments on AMT, one of the biggest and most popular crowdsourcing platforms.
While AMT is similar to many other crowdsourcing platforms, and while we do use a custom annotation platform which limits annotators interaction with AMT-specific UI, there are a few unique traits to AMT.
First, the population of workers is hard to replicate to other platforms.
We use several of AMT's built in qualifications to filter out workers, and there is no clear translation for which qualifications to use on other crowdsourcing platforms.
Further, AMT has different payment expectations than other crowdsourcing communities.
Some crowdsourcing platforms are purely volunteer based, while others attract short-time workers who complete only a few tasks.
Ultimately, our choice to work on AMT is motivated by the size and popularity of this platform, thereby having results be relevant for a large set of researchers.

\textit{Omission of binary-label formats}.
Due to cost constraints, we could not experiment with labeling schemes which involved presenting binary choices to the workers.
While this is representative of real-world scenarios for tasks similar to ours, it also leaves questions regarding whether or not the quality of annotation is potentially higher with these methods.
However, we believe that given the vast cost differences of these methods, this choice is a reasonable assumption and will closely represent decisions made by the researchers for whom this analysis is intended.

\textit{Generalizing to new hierarchies}.
While we have no clear reason to expect our results are specific to the vaccine concerns hierarchy we used, we do not show or indicate that these results generalize well beyond it.
For instance, as the size of the hierarchy grows, one might expect that the single-pass options become cognitively overbearing, and therefore multi-pass methods might begin increasing in relative performance.
However, in offering useful analysis this is a choice that must be made.

\new{
\textit{Budget implications}.
We set a single budget and examine how to best optimize annotation performance against ground-truth labels on AMT. However, it may be the case that the best labeling scheme for our budget shows a less significant improvement when the budget is much higher and the reward given to workers is increased. Or there may be a different labeling scheme which performs better under a higher budget. One could imagine repeating our experiments at several budgets, and examining the relationship between a particular labeling scheme's data quality and the relative expense of data collection. Some schemes may be more cost efficient, showing small differences in worker performance across budgets, while others may only become viable at higher budgets. 
Unfortunately, running such experiments would make this research prohibitively expensive.
}
\new{
The budget can be set 
based on previous experimentation regarding the minimum budget needed to achieve relatively high quality data, as well as confidently exceed the United States minimum wage. Teams that wish to collect data would likely avoid opting to pay more for the labels they are getting. For machine learning applications, it is well known that you may get more utility from collecting \textit{additional} data, rather than increasing the quality of the labeled data \cite{halevy2009unreasonable}.
}

\section{Conclusion}
\label{conclusion}

We investigate various labeling schemes for crowdsourcing text annotation of difficult, high-subjectivity tasks and measure impact on worker performance against ground-truth labels.
We find that integrating hierarchies into the labeling scheme helps with boosting performance.

\new{Through analysis, we explore three potential indirect causes for improvement against ground-truth labels:} 
    (1) They group similar concepts together, improving F1 scores to $0.50$ from $0.34$ as compared to random groupings. 
    (2) They allow relative increases in performance on difficult passages, leading to an increase in as much as $+0.40$ on F1 score on high difficulty examples.
    (3) They boost the true positive frequency, thereby increasing precision of workers without detriment to recall.
We recommend considering incorporating hierarchies into the labeling process if optimizing for individual worker performance, while using a majority vote setting if solely optimizing for F1 score (achieving $0.70$ with \textit{single-pass hierarchical multi-label}).

\begin{acks}
We thank Pardis Emami-Naeini and anonymous reviewers for feedback.
This work was supported by NSF award IIS-2211526 and an
award from Google.
\end{acks}

\bibliographystyle{ACM-Reference-Format}
\bibliography{bibs/bibliography}


\begin{thebibliography}{58}


\ifx \showCODEN    \undefined \def \showCODEN     #1{\unskip}     \fi
\ifx \showDOI      \undefined \def \showDOI       #1{#1}\fi
\ifx \showISBNx    \undefined \def \showISBNx     #1{\unskip}     \fi
\ifx \showISBNxiii \undefined \def \showISBNxiii  #1{\unskip}     \fi
\ifx \showISSN     \undefined \def \showISSN      #1{\unskip}     \fi
\ifx \showLCCN     \undefined \def \showLCCN      #1{\unskip}     \fi
\ifx \shownote     \undefined \def \shownote      #1{#1}          \fi
\ifx \showarticletitle \undefined \def \showarticletitle #1{#1}   \fi
\ifx \showURL      \undefined \def \showURL       {\relax}        \fi
\providecommand\bibfield[2]{#2}
\providecommand\bibinfo[2]{#2}
\providecommand\natexlab[1]{#1}
\providecommand\showeprint[2][]{arXiv:#2}

\bibitem[Alemanno(2018)]%
        {alemanno2018counter}
\bibfield{author}{\bibinfo{person}{Alberto Alemanno}.}
  \bibinfo{year}{2018}\natexlab{}.
\newblock \showarticletitle{How to counter fake news? A taxonomy of anti-fake
  news approaches}.
\newblock \bibinfo{journal}{\emph{European journal of risk regulation}}
  \bibinfo{volume}{9}, \bibinfo{number}{1} (\bibinfo{year}{2018}),
  \bibinfo{pages}{1--5}.
\newblock


\bibitem[Arslan et~al\mbox{.}(2020a)]%
        {arslan_modeling_nodate}
\bibfield{author}{\bibinfo{person}{Fatma Arslan}, \bibinfo{person}{Josue
  Caraballo}, \bibinfo{person}{Damian Jimenez}, {and} \bibinfo{person}{Chengkai
  Li}.} \bibinfo{year}{2020}\natexlab{a}.
\newblock \showarticletitle{Modeling Factual Claims with Semantic Frames}. In
  \bibinfo{booktitle}{\emph{Proceedings of the Twelfth Language Resources and
  Evaluation Conference}}. \bibinfo{publisher}{European Language Resources
  Association}, \bibinfo{address}{Marseille, France},
  \bibinfo{pages}{2511--2520}.
\newblock
\showISBNx{979-10-95546-34-4}
\urldef\tempurl%
\url{https://aclanthology.org/2020.lrec-1.306}
\showURL{%
\tempurl}


\bibitem[Arslan et~al\mbox{.}(2020b)]%
        {arslan_benchmark_2020}
\bibfield{author}{\bibinfo{person}{Fatma Arslan}, \bibinfo{person}{Naeemul
  Hassan}, \bibinfo{person}{Chengkai Li}, {and} \bibinfo{person}{Mark
  Tremayne}.} \bibinfo{year}{2020}\natexlab{b}.
\newblock \showarticletitle{A {Benchmark} {Dataset} of {Check}-{Worthy}
  {Factual} {Claims}}.
\newblock \bibinfo{journal}{\emph{Proceedings of the International AAAI
  Conference on Web and Social Media}}  \bibinfo{volume}{14}
  (\bibinfo{date}{May} \bibinfo{year}{2020}), \bibinfo{pages}{821--829}.
\newblock
\showISSN{2334-0770}
\urldef\tempurl%
\url{https://ojs.aaai.org/index.php/ICWSM/article/view/7346}
\showURL{%
\tempurl}


\bibitem[Barbosa and Chen(2019)]%
        {barbosa_rehumanized_2019}
\bibfield{author}{\bibinfo{person}{Natã~M. Barbosa} {and}
  \bibinfo{person}{Monchu Chen}.} \bibinfo{year}{2019}\natexlab{}.
\newblock \showarticletitle{Rehumanized {Crowdsourcing}: {A} {Labeling}
  {Framework} {Addressing} {Bias} and {Ethics} in {Machine} {Learning}}. In
  \bibinfo{booktitle}{\emph{Proceedings of the 2019 {CHI} {Conference} on
  {Human} {Factors} in {Computing} {Systems}}} \emph{(\bibinfo{series}{{CHI}
  '19})}. \bibinfo{publisher}{Association for Computing Machinery},
  \bibinfo{address}{New York, NY, USA}, \bibinfo{pages}{1--12}.
\newblock
\showISBNx{978-1-4503-5970-2}
\urldef\tempurl%
\url{https://doi.org/10.1145/3290605.3300773}
\showDOI{\tempurl}


\bibitem[Bowman et~al\mbox{.}(2015)]%
        {bowman_snli_2015}
\bibfield{author}{\bibinfo{person}{Samuel~R. Bowman}, \bibinfo{person}{Gabor
  Angeli}, \bibinfo{person}{Christopher Potts}, {and}
  \bibinfo{person}{Christopher~D. Manning}.} \bibinfo{year}{2015}\natexlab{}.
\newblock \bibinfo{title}{A large annotated corpus for learning natural
  language inference}.
\newblock
\newblock
\urldef\tempurl%
\url{https://doi.org/10.48550/arXiv.1508.05326}
\showDOI{\tempurl}
\newblock
\shownote{arXiv:1508.05326 [cs]}.


\bibitem[Bragg et~al\mbox{.}(2013)]%
        {bragg_crowdsourcing_2013}
\bibfield{author}{\bibinfo{person}{Jonathan Bragg}, \bibinfo{person}{Mausam},
  {and} \bibinfo{person}{Daniel Weld}.} \bibinfo{year}{2013}\natexlab{}.
\newblock \showarticletitle{Crowdsourcing {Multi}-{Label} {Classification} for
  {Taxonomy} {Creation}}.
\newblock \bibinfo{journal}{\emph{Proceedings of the AAAI Conference on Human
  Computation and Crowdsourcing}}  \bibinfo{volume}{1} (\bibinfo{date}{Nov.}
  \bibinfo{year}{2013}), \bibinfo{pages}{25--33}.
\newblock
\showISSN{2769-1349}
\urldef\tempurl%
\url{https://ojs.aaai.org/index.php/HCOMP/article/view/13091}
\showURL{%
\tempurl}


\bibitem[Cartwright et~al\mbox{.}(2019)]%
        {cartwright_audio_2019}
\bibfield{author}{\bibinfo{person}{Mark Cartwright}, \bibinfo{person}{Graham
  Dove}, \bibinfo{person}{Ana~Elisa Méndez~Méndez}, \bibinfo{person}{Juan~P.
  Bello}, {and} \bibinfo{person}{Oded Nov}.} \bibinfo{year}{2019}\natexlab{}.
\newblock \showarticletitle{Crowdsourcing {Multi}-label {Audio} {Annotation}
  {Tasks} with {Citizen} {Scientists}}. In
  \bibinfo{booktitle}{\emph{Proceedings of the 2019 {CHI} {Conference} on
  {Human} {Factors} in {Computing} {Systems}}} \emph{(\bibinfo{series}{{CHI}
  '19})}. \bibinfo{publisher}{Association for Computing Machinery},
  \bibinfo{address}{New York, NY, USA}, \bibinfo{pages}{1--11}.
\newblock
\showISBNx{978-1-4503-5970-2}
\urldef\tempurl%
\url{https://doi.org/10.1145/3290605.3300522}
\showDOI{\tempurl}


\bibitem[Chen et~al\mbox{.}(2010)]%
        {chen_efficient_2010}
\bibfield{author}{\bibinfo{person}{Xiangyu Chen}, \bibinfo{person}{Yadong Mu},
  \bibinfo{person}{Shuicheng Yan}, {and} \bibinfo{person}{Tat-Seng Chua}.}
  \bibinfo{year}{2010}\natexlab{}.
\newblock \showarticletitle{Efficient large-scale image annotation by
  probabilistic collaborative multi-label propagation}. In
  \bibinfo{booktitle}{\emph{Proceedings of the 18th {ACM} international
  conference on {Multimedia}}} \emph{(\bibinfo{series}{{MM} '10})}.
  \bibinfo{publisher}{Association for Computing Machinery},
  \bibinfo{address}{New York, NY, USA}, \bibinfo{pages}{35--44}.
\newblock
\showISBNx{978-1-60558-933-6}
\urldef\tempurl%
\url{https://doi.org/10.1145/1873951.1873959}
\showDOI{\tempurl}


\bibitem[Chilton et~al\mbox{.}(2013)]%
        {chilton_cascade_2013}
\bibfield{author}{\bibinfo{person}{Lydia~B. Chilton}, \bibinfo{person}{Greg
  Little}, \bibinfo{person}{Darren Edge}, \bibinfo{person}{Daniel~S. Weld},
  {and} \bibinfo{person}{James~A. Landay}.} \bibinfo{year}{2013}\natexlab{}.
\newblock \showarticletitle{Cascade: crowdsourcing taxonomy creation}. In
  \bibinfo{booktitle}{\emph{Proceedings of the {SIGCHI} {Conference} on {Human}
  {Factors} in {Computing} {Systems}}} \emph{(\bibinfo{series}{{CHI} '13})}.
  \bibinfo{publisher}{Association for Computing Machinery},
  \bibinfo{address}{New York, NY, USA}, \bibinfo{pages}{1999--2008}.
\newblock
\showISBNx{978-1-4503-1899-0}
\urldef\tempurl%
\url{https://doi.org/10.1145/2470654.2466265}
\showDOI{\tempurl}


\bibitem[Coan et~al\mbox{.}(2021)]%
        {coan_climate_2021}
\bibfield{author}{\bibinfo{person}{Travis~G. Coan},
  \bibinfo{person}{Constantine Boussalis}, \bibinfo{person}{John Cook}, {and}
  \bibinfo{person}{Mirjam~O. Nanko}.} \bibinfo{year}{2021}\natexlab{}.
\newblock \showarticletitle{Computer-assisted classification of contrarian
  claims about climate change}.
\newblock \bibinfo{journal}{\emph{Scientific Reports}} \bibinfo{volume}{11},
  \bibinfo{number}{1} (\bibinfo{date}{Nov.} \bibinfo{year}{2021}),
  \bibinfo{pages}{22320}.
\newblock
\showISSN{2045-2322}
\urldef\tempurl%
\url{https://doi.org/10.1038/s41598-021-01714-4}
\showDOI{\tempurl}
\newblock
\shownote{Number: 1 Publisher: Nature Publishing Group}.


\bibitem[Dawid and Skene(1979)]%
        {dawid_maximum_1979}
\bibfield{author}{\bibinfo{person}{A.~P. Dawid} {and} \bibinfo{person}{A.~M.
  Skene}.} \bibinfo{year}{1979}\natexlab{}.
\newblock \showarticletitle{Maximum {Likelihood} {Estimation} of {Observer}
  {Error}-{Rates} {Using} the {EM} {Algorithm}}.
\newblock \bibinfo{journal}{\emph{Applied Statistics}} \bibinfo{volume}{28},
  \bibinfo{number}{1} (\bibinfo{year}{1979}), \bibinfo{pages}{20}.
\newblock
\showISSN{00359254}
\urldef\tempurl%
\url{https://doi.org/10.2307/2346806}
\showDOI{\tempurl}


\bibitem[Defferrard et~al\mbox{.}(2017)]%
        {defferrard_fma_2017}
\bibfield{author}{\bibinfo{person}{Michaël Defferrard},
  \bibinfo{person}{Kirell Benzi}, \bibinfo{person}{Pierre Vandergheynst}, {and}
  \bibinfo{person}{Xavier Bresson}.} \bibinfo{year}{2017}\natexlab{}.
\newblock \bibinfo{title}{{FMA}: {A} {Dataset} {For} {Music} {Analysis}}.
\newblock
\newblock
\urldef\tempurl%
\url{https://doi.org/10.48550/arXiv.1612.01840}
\showDOI{\tempurl}
\newblock
\shownote{arXiv:1612.01840 [cs]}.


\bibitem[Deng et~al\mbox{.}(2009)]%
        {imagenet}
\bibfield{author}{\bibinfo{person}{Jia Deng}, \bibinfo{person}{Wei Dong},
  \bibinfo{person}{Richard Socher}, \bibinfo{person}{Li-Jia Li},
  \bibinfo{person}{Kai Li}, {and} \bibinfo{person}{Li Fei-Fei}.}
  \bibinfo{year}{2009}\natexlab{}.
\newblock \showarticletitle{ImageNet: A large-scale hierarchical image
  database}. In \bibinfo{booktitle}{\emph{2009 IEEE Conference on Computer
  Vision and Pattern Recognition}}. \bibinfo{pages}{248--255}.
\newblock
\urldef\tempurl%
\url{https://doi.org/10.1109/CVPR.2009.5206848}
\showDOI{\tempurl}


\bibitem[Deng et~al\mbox{.}(2014)]%
        {deng_scalable_2014}
\bibfield{author}{\bibinfo{person}{Jia Deng}, \bibinfo{person}{Olga
  Russakovsky}, \bibinfo{person}{Jonathan Krause}, \bibinfo{person}{Michael~S.
  Bernstein}, \bibinfo{person}{Alex Berg}, {and} \bibinfo{person}{Li Fei-Fei}.}
  \bibinfo{year}{2014}\natexlab{}.
\newblock \showarticletitle{Scalable multi-label annotation}. In
  \bibinfo{booktitle}{\emph{Proceedings of the {SIGCHI} {Conference} on {Human}
  {Factors} in {Computing} {Systems}}} \emph{(\bibinfo{series}{{CHI} '14})}.
  \bibinfo{publisher}{Association for Computing Machinery},
  \bibinfo{address}{New York, NY, USA}, \bibinfo{pages}{3099--3102}.
\newblock
\showISBNx{978-1-4503-2473-1}
\urldef\tempurl%
\url{https://doi.org/10.1145/2556288.2557011}
\showDOI{\tempurl}


\bibitem[Eveleigh et~al\mbox{.}(2014)]%
        {eveleigh_dabblers_2014}
\bibfield{author}{\bibinfo{person}{Alexandra Eveleigh},
  \bibinfo{person}{Charlene Jennett}, \bibinfo{person}{Ann Blandford},
  \bibinfo{person}{Philip Brohan}, {and} \bibinfo{person}{Anna~L. Cox}.}
  \bibinfo{year}{2014}\natexlab{}.
\newblock \showarticletitle{Designing for dabblers and deterring drop-outs in
  citizen science}. In \bibinfo{booktitle}{\emph{Proceedings of the {SIGCHI}
  {Conference} on {Human} {Factors} in {Computing} {Systems}}}
  \emph{(\bibinfo{series}{{CHI} '14})}. \bibinfo{publisher}{Association for
  Computing Machinery}, \bibinfo{address}{New York, NY, USA},
  \bibinfo{pages}{2985--2994}.
\newblock
\showISBNx{978-1-4503-2473-1}
\urldef\tempurl%
\url{https://doi.org/10.1145/2556288.2557262}
\showDOI{\tempurl}


\bibitem[Findlater et~al\mbox{.}(2017)]%
        {findlater_differences_2017}
\bibfield{author}{\bibinfo{person}{Leah Findlater}, \bibinfo{person}{Joan
  Zhang}, \bibinfo{person}{Jon~E. Froehlich}, {and} \bibinfo{person}{Karyn
  Moffatt}.} \bibinfo{year}{2017}\natexlab{}.
\newblock \showarticletitle{Differences in {Crowdsourced} vs. {Lab}-based
  {Mobile} and {Desktop} {Input} {Performance} {Data}}. In
  \bibinfo{booktitle}{\emph{Proceedings of the 2017 {CHI} {Conference} on
  {Human} {Factors} in {Computing} {Systems}}} \emph{(\bibinfo{series}{{CHI}
  '17})}. \bibinfo{publisher}{Association for Computing Machinery},
  \bibinfo{address}{New York, NY, USA}, \bibinfo{pages}{6813--6824}.
\newblock
\showISBNx{978-1-4503-4655-9}
\urldef\tempurl%
\url{https://doi.org/10.1145/3025453.3025820}
\showDOI{\tempurl}


\bibitem[for Computing~Machinery(2023)]%
        {acm_ccs}
\bibfield{author}{\bibinfo{person}{Association for Computing~Machinery}.}
  \bibinfo{year}{2023}\natexlab{}.
\newblock \bibinfo{title}{The 2012 {ACM} {Computing} {Classification}
  {System}}.
\newblock
\newblock
\urldef\tempurl%
\url{https://www.acm.org/publications/class-2012}
\showURL{%
\tempurl}


\bibitem[Gemmeke et~al\mbox{.}(2017)]%
        {gemmeke_audio_2017}
\bibfield{author}{\bibinfo{person}{Jort~F. Gemmeke}, \bibinfo{person}{Daniel
  P.~W. Ellis}, \bibinfo{person}{Dylan Freedman}, \bibinfo{person}{Aren
  Jansen}, \bibinfo{person}{Wade Lawrence}, \bibinfo{person}{R.~Channing
  Moore}, \bibinfo{person}{Manoj Plakal}, {and} \bibinfo{person}{Marvin
  Ritter}.} \bibinfo{year}{2017}\natexlab{}.
\newblock \showarticletitle{Audio {Set}: {An} ontology and human-labeled
  dataset for audio events}. In \bibinfo{booktitle}{\emph{2017 {IEEE}
  {International} {Conference} on {Acoustics}, {Speech} and {Signal}
  {Processing} ({ICASSP})}}. \bibinfo{pages}{776--780}.
\newblock
\urldef\tempurl%
\url{https://doi.org/10.1109/ICASSP.2017.7952261}
\showDOI{\tempurl}
\newblock
\shownote{ISSN: 2379-190X}.


\bibitem[George~Saadé and Alexandre~Otrakji(2007)]%
        {george_saade_first_2007}
\bibfield{author}{\bibinfo{person}{Raafat George~Saadé} {and}
  \bibinfo{person}{Camille Alexandre~Otrakji}.}
  \bibinfo{year}{2007}\natexlab{}.
\newblock \showarticletitle{First impressions last a lifetime: effect of
  interface type on disorientation and cognitive load}.
\newblock \bibinfo{journal}{\emph{Computers in Human Behavior}}
  \bibinfo{volume}{23}, \bibinfo{number}{1} (\bibinfo{date}{Jan.}
  \bibinfo{year}{2007}), \bibinfo{pages}{525--535}.
\newblock
\showISSN{0747-5632}
\urldef\tempurl%
\url{https://doi.org/10.1016/j.chb.2004.10.035}
\showDOI{\tempurl}


\bibitem[Halevy et~al\mbox{.}(2009)]%
        {halevy2009unreasonable}
\bibfield{author}{\bibinfo{person}{Alon Halevy}, \bibinfo{person}{Peter
  Norvig}, {and} \bibinfo{person}{Fernando Pereira}.}
  \bibinfo{year}{2009}\natexlab{}.
\newblock \showarticletitle{The unreasonable effectiveness of data}.
\newblock \bibinfo{journal}{\emph{IEEE intelligent systems}}
  \bibinfo{volume}{24}, \bibinfo{number}{2} (\bibinfo{year}{2009}),
  \bibinfo{pages}{8--12}.
\newblock


\bibitem[Humphrey et~al\mbox{.}(2018)]%
        {humphrey_openmic_2018}
\bibfield{author}{\bibinfo{person}{Eric Humphrey}, \bibinfo{person}{Simon
  Durand}, {and} \bibinfo{person}{Brian McFee}.}
  \bibinfo{year}{2018}\natexlab{}.
\newblock \showarticletitle{OpenMIC-2018: An Open Data-set for Multiple
  Instrument Recognition.}. In \bibinfo{booktitle}{\emph{ISMIR}}.
  \bibinfo{pages}{438--444}.
\newblock


\bibitem[Iacovides et~al\mbox{.}(2013)]%
        {iacovides_games_2013}
\bibfield{author}{\bibinfo{person}{Ioanna Iacovides}, \bibinfo{person}{Charlene
  Jennett}, \bibinfo{person}{Cassandra Cornish-Trestrail}, {and}
  \bibinfo{person}{Anna~L. Cox}.} \bibinfo{year}{2013}\natexlab{}.
\newblock \showarticletitle{Do games attract or sustain engagement in citizen
  science? a study of volunteer motivations}. In
  \bibinfo{booktitle}{\emph{{CHI} '13 {Extended} {Abstracts} on {Human}
  {Factors} in {Computing} {Systems}}} \emph{(\bibinfo{series}{{CHI} {EA}
  '13})}. \bibinfo{publisher}{Association for Computing Machinery},
  \bibinfo{address}{New York, NY, USA}, \bibinfo{pages}{1101--1106}.
\newblock
\showISBNx{978-1-4503-1952-2}
\urldef\tempurl%
\url{https://doi.org/10.1145/2468356.2468553}
\showDOI{\tempurl}


\bibitem[Irani and Silberman(2013)]%
        {irani_turkopticon_2013}
\bibfield{author}{\bibinfo{person}{Lilly~C. Irani} {and}
  \bibinfo{person}{M.~Six Silberman}.} \bibinfo{year}{2013}\natexlab{}.
\newblock \showarticletitle{Turkopticon: interrupting worker invisibility in
  amazon mechanical turk}. In \bibinfo{booktitle}{\emph{Proceedings of the
  {SIGCHI} {Conference} on {Human} {Factors} in {Computing} {Systems}}}
  \emph{(\bibinfo{series}{{CHI} '13})}. \bibinfo{publisher}{Association for
  Computing Machinery}, \bibinfo{address}{New York, NY, USA},
  \bibinfo{pages}{611--620}.
\newblock
\showISBNx{978-1-4503-1899-0}
\urldef\tempurl%
\url{https://doi.org/10.1145/2470654.2470742}
\showDOI{\tempurl}


\bibitem[Jacobson et~al\mbox{.}(2007)]%
        {jacobson_taxonomy_2007}
\bibfield{author}{\bibinfo{person}{Robert~M. Jacobson},
  \bibinfo{person}{Paul~V. Targonski}, {and} \bibinfo{person}{Gregory~A.
  Poland}.} \bibinfo{year}{2007}\natexlab{}.
\newblock \showarticletitle{A taxonomy of reasoning flaws in the anti-vaccine
  movement}.
\newblock \bibinfo{journal}{\emph{Vaccine}} \bibinfo{volume}{25},
  \bibinfo{number}{16} (\bibinfo{date}{April} \bibinfo{year}{2007}),
  \bibinfo{pages}{3146--3152}.
\newblock
\showISSN{0264-410X}
\urldef\tempurl%
\url{https://doi.org/10.1016/j.vaccine.2007.01.046}
\showDOI{\tempurl}


\bibitem[Kay et~al\mbox{.}(2017)]%
        {kay_kinetics_2017}
\bibfield{author}{\bibinfo{person}{Will Kay}, \bibinfo{person}{Joao Carreira},
  \bibinfo{person}{Karen Simonyan}, \bibinfo{person}{Brian Zhang},
  \bibinfo{person}{Chloe Hillier}, \bibinfo{person}{Sudheendra
  Vijayanarasimhan}, \bibinfo{person}{Fabio Viola}, \bibinfo{person}{Tim
  Green}, \bibinfo{person}{Trevor Back}, \bibinfo{person}{Paul Natsev},
  \bibinfo{person}{Mustafa Suleyman}, {and} \bibinfo{person}{Andrew
  Zisserman}.} \bibinfo{year}{2017}\natexlab{}.
\newblock \bibinfo{title}{The {Kinetics} {Human} {Action} {Video} {Dataset}}.
\newblock
\newblock
\urldef\tempurl%
\url{https://doi.org/10.48550/arXiv.1705.06950}
\showDOI{\tempurl}
\newblock
\shownote{arXiv:1705.06950 [cs]}.


\bibitem[Khetan and Oh(2017)]%
        {khetan_achieving_2017}
\bibfield{author}{\bibinfo{person}{Ashish Khetan} {and}
  \bibinfo{person}{Sewoong Oh}.} \bibinfo{year}{2017}\natexlab{}.
\newblock \bibinfo{title}{Achieving {Budget}-optimality with {Adaptive}
  {Schemes} in {Crowdsourcing}}.
\newblock
\newblock
\urldef\tempurl%
\url{http://arxiv.org/abs/1602.03481}
\showURL{%
\tempurl}
\newblock
\shownote{arXiv:1602.03481 [cs, stat]}.


\bibitem[Kittur et~al\mbox{.}(2008)]%
        {kittur_userstudies_2008}
\bibfield{author}{\bibinfo{person}{Aniket Kittur}, \bibinfo{person}{Ed~H. Chi},
  {and} \bibinfo{person}{Bongwon Suh}.} \bibinfo{year}{2008}\natexlab{}.
\newblock \showarticletitle{Crowdsourcing user studies with {Mechanical}
  {Turk}}. In \bibinfo{booktitle}{\emph{Proceedings of the {SIGCHI}
  {Conference} on {Human} {Factors} in {Computing} {Systems}}}
  \emph{(\bibinfo{series}{{CHI} '08})}. \bibinfo{publisher}{Association for
  Computing Machinery}, \bibinfo{address}{New York, NY, USA},
  \bibinfo{pages}{453--456}.
\newblock
\showISBNx{978-1-60558-011-1}
\urldef\tempurl%
\url{https://doi.org/10.1145/1357054.1357127}
\showDOI{\tempurl}


\bibitem[Kittur et~al\mbox{.}(2011)]%
        {kittur_crowdforge_2011}
\bibfield{author}{\bibinfo{person}{Aniket Kittur}, \bibinfo{person}{Boris
  Smus}, {and} \bibinfo{person}{Robert Kraut}.}
  \bibinfo{year}{2011}\natexlab{}.
\newblock \showarticletitle{{CrowdForge}: crowdsourcing complex work}. In
  \bibinfo{booktitle}{\emph{{CHI} '11 {Extended} {Abstracts} on {Human}
  {Factors} in {Computing} {Systems}}} \emph{(\bibinfo{series}{{CHI} {EA}
  '11})}. \bibinfo{publisher}{Association for Computing Machinery},
  \bibinfo{address}{New York, NY, USA}, \bibinfo{pages}{1801--1806}.
\newblock
\showISBNx{978-1-4503-0268-5}
\urldef\tempurl%
\url{https://doi.org/10.1145/1979742.1979902}
\showDOI{\tempurl}


\bibitem[Krishna et~al\mbox{.}(2016)]%
        {krishna_embracing_2016}
\bibfield{author}{\bibinfo{person}{Ranjay Krishna}, \bibinfo{person}{Kenji
  Hata}, \bibinfo{person}{Stephanie Chen}, \bibinfo{person}{Joshua Kravitz},
  \bibinfo{person}{David~A. Shamma}, \bibinfo{person}{Li Fei-Fei}, {and}
  \bibinfo{person}{Michael~S. Bernstein}.} \bibinfo{year}{2016}\natexlab{}.
\newblock \showarticletitle{Embracing {Error} to {Enable} {Rapid}
  {Crowdsourcing}}. In \bibinfo{booktitle}{\emph{Proceedings of the 2016 {CHI}
  {Conference} on {Human} {Factors} in {Computing} {Systems}}}.
  \bibinfo{pages}{3167--3179}.
\newblock
\urldef\tempurl%
\url{https://doi.org/10.1145/2858036.2858115}
\showDOI{\tempurl}
\newblock
\shownote{arXiv:1602.04506 [cs]}.


\bibitem[Kuznetsova et~al\mbox{.}(2020)]%
        {kuznetsova_open_2020}
\bibfield{author}{\bibinfo{person}{Alina Kuznetsova}, \bibinfo{person}{Hassan
  Rom}, \bibinfo{person}{Neil Alldrin}, \bibinfo{person}{Jasper Uijlings},
  \bibinfo{person}{Ivan Krasin}, \bibinfo{person}{Jordi Pont-Tuset},
  \bibinfo{person}{Shahab Kamali}, \bibinfo{person}{Stefan Popov},
  \bibinfo{person}{Matteo Malloci}, \bibinfo{person}{Alexander Kolesnikov},
  \bibinfo{person}{Tom Duerig}, {and} \bibinfo{person}{Vittorio Ferrari}.}
  \bibinfo{year}{2020}\natexlab{}.
\newblock \showarticletitle{The {Open} {Images} {Dataset} {V4}}.
\newblock \bibinfo{journal}{\emph{International Journal of Computer Vision}}
  \bibinfo{volume}{128}, \bibinfo{number}{7} (\bibinfo{date}{July}
  \bibinfo{year}{2020}), \bibinfo{pages}{1956--1981}.
\newblock
\showISSN{1573-1405}
\urldef\tempurl%
\url{https://doi.org/10.1007/s11263-020-01316-z}
\showDOI{\tempurl}


\bibitem[Lee et~al\mbox{.}(2013)]%
        {lee_greenify_2013}
\bibfield{author}{\bibinfo{person}{Joey~J. Lee}, \bibinfo{person}{Eduard
  Matamoros}, \bibinfo{person}{Rafael Kern}, \bibinfo{person}{Jenna Marks},
  \bibinfo{person}{Christian de Luna}, {and} \bibinfo{person}{William
  Jordan-Cooley}.} \bibinfo{year}{2013}\natexlab{}.
\newblock \showarticletitle{Greenify: fostering sustainable communities via
  gamification}. In \bibinfo{booktitle}{\emph{{CHI} '13 {Extended} {Abstracts}
  on {Human} {Factors} in {Computing} {Systems}}} \emph{(\bibinfo{series}{{CHI}
  {EA} '13})}. \bibinfo{publisher}{Association for Computing Machinery},
  \bibinfo{address}{New York, NY, USA}, \bibinfo{pages}{1497--1502}.
\newblock
\showISBNx{978-1-4503-1952-2}
\urldef\tempurl%
\url{https://doi.org/10.1145/2468356.2468623}
\showDOI{\tempurl}


\bibitem[Lin et~al\mbox{.}(2015)]%
        {lin_microsoft_2015}
\bibfield{author}{\bibinfo{person}{Tsung-Yi Lin}, \bibinfo{person}{Michael
  Maire}, \bibinfo{person}{Serge Belongie}, \bibinfo{person}{Lubomir Bourdev},
  \bibinfo{person}{Ross Girshick}, \bibinfo{person}{James Hays},
  \bibinfo{person}{Pietro Perona}, \bibinfo{person}{Deva Ramanan},
  \bibinfo{person}{C.~Lawrence Zitnick}, {and} \bibinfo{person}{Piotr
  Dollár}.} \bibinfo{year}{2015}\natexlab{}.
\newblock \bibinfo{title}{Microsoft {COCO}: {Common} {Objects} in {Context}}.
\newblock
\newblock
\urldef\tempurl%
\url{https://doi.org/10.48550/arXiv.1405.0312}
\showDOI{\tempurl}
\newblock
\shownote{arXiv:1405.0312 [cs]}.


\bibitem[Lintott et~al\mbox{.}(2008)]%
        {lintott_galaxy_2008}
\bibfield{author}{\bibinfo{person}{Chris~J. Lintott}, \bibinfo{person}{Kevin
  Schawinski}, \bibinfo{person}{Anze Slosar}, \bibinfo{person}{Kate Land},
  \bibinfo{person}{Steven Bamford}, \bibinfo{person}{Daniel Thomas},
  \bibinfo{person}{M.~Jordan Raddick}, \bibinfo{person}{Robert~C. Nichol},
  \bibinfo{person}{Alex Szalay}, \bibinfo{person}{Dan Andreescu},
  \bibinfo{person}{Phil Murray}, {and} \bibinfo{person}{Jan van~den Berg}.}
  \bibinfo{year}{2008}\natexlab{}.
\newblock \showarticletitle{Galaxy {Zoo} : {Morphologies} derived from visual
  inspection of galaxies from the {Sloan} {Digital} {Sky} {Survey}}.
\newblock \bibinfo{journal}{\emph{Monthly Notices of the Royal Astronomical
  Society}} \bibinfo{volume}{389}, \bibinfo{number}{3} (\bibinfo{date}{Sept.}
  \bibinfo{year}{2008}), \bibinfo{pages}{1179--1189}.
\newblock
\showISSN{00358711, 13652966}
\urldef\tempurl%
\url{https://doi.org/10.1111/j.1365-2966.2008.13689.x}
\showDOI{\tempurl}
\newblock
\shownote{arXiv:0804.4483 [astro-ph]}.


\bibitem[Liu et~al\mbox{.}(2015)]%
        {celeb_faces}
\bibfield{author}{\bibinfo{person}{Ziwei Liu}, \bibinfo{person}{Ping Luo},
  \bibinfo{person}{Xiaogang Wang}, {and} \bibinfo{person}{Xiaoou Tang}.}
  \bibinfo{year}{2015}\natexlab{}.
\newblock \showarticletitle{Deep {Learning} {Face} {Attributes} in the {Wild}}.
  In \bibinfo{booktitle}{\emph{2015 {IEEE} {International} {Conference} on
  {Computer} {Vision} ({ICCV})}}. \bibinfo{pages}{3730--3738}.
\newblock
\urldef\tempurl%
\url{https://doi.org/10.1109/ICCV.2015.425}
\showDOI{\tempurl}
\newblock
\shownote{ISSN: 2380-7504}.


\bibitem[MacKenzie and Ware(1993)]%
        {mackenzie_lag_1993}
\bibfield{author}{\bibinfo{person}{I.~Scott MacKenzie} {and}
  \bibinfo{person}{Colin Ware}.} \bibinfo{year}{1993}\natexlab{}.
\newblock \showarticletitle{Lag as a determinant of human performance in
  interactive systems}. In \bibinfo{booktitle}{\emph{Proceedings of the
  {INTERACT} '93 and {CHI} '93 {Conference} on {Human} {Factors} in {Computing}
  {Systems}}} \emph{(\bibinfo{series}{{CHI} '93})}.
  \bibinfo{publisher}{Association for Computing Machinery},
  \bibinfo{address}{New York, NY, USA}, \bibinfo{pages}{488--493}.
\newblock
\showISBNx{978-0-89791-575-5}
\urldef\tempurl%
\url{https://doi.org/10.1145/169059.169431}
\showDOI{\tempurl}


\bibitem[Miyata et~al\mbox{.}(2022)]%
        {miyata_gamification_2022}
\bibfield{author}{\bibinfo{person}{Akihiro Miyata}, \bibinfo{person}{Yusaku
  Murayama}, \bibinfo{person}{Akihiro Furuta}, \bibinfo{person}{Kazuki
  Okugawa}, \bibinfo{person}{Keihiro Ochiai}, {and} \bibinfo{person}{Yuko
  Murayama}.} \bibinfo{year}{2022}\natexlab{}.
\newblock \showarticletitle{Gamification strategies to improve the motivation
  and performance in accessibility information collection}. In
  \bibinfo{booktitle}{\emph{Extended {Abstracts} of the 2022 {CHI} {Conference}
  on {Human} {Factors} in {Computing} {Systems}}} \emph{(\bibinfo{series}{{CHI}
  {EA} '22})}. \bibinfo{publisher}{Association for Computing Machinery},
  \bibinfo{address}{New York, NY, USA}, \bibinfo{pages}{1--7}.
\newblock
\showISBNx{978-1-4503-9156-6}
\urldef\tempurl%
\url{https://doi.org/10.1145/3491101.3519783}
\showDOI{\tempurl}


\bibitem[Morris et~al\mbox{.}(2017)]%
        {morris_subcontracting_2017}
\bibfield{author}{\bibinfo{person}{Meredith~Ringel Morris},
  \bibinfo{person}{Jeffrey~P. Bigham}, \bibinfo{person}{Robin Brewer},
  \bibinfo{person}{Jonathan Bragg}, \bibinfo{person}{Anand Kulkarni},
  \bibinfo{person}{Jessie Li}, {and} \bibinfo{person}{Saiph Savage}.}
  \bibinfo{year}{2017}\natexlab{}.
\newblock \showarticletitle{Subcontracting {Microwork}}. In
  \bibinfo{booktitle}{\emph{Proceedings of the 2017 {CHI} {Conference} on
  {Human} {Factors} in {Computing} {Systems}}} \emph{(\bibinfo{series}{{CHI}
  '17})}. \bibinfo{publisher}{Association for Computing Machinery},
  \bibinfo{address}{New York, NY, USA}, \bibinfo{pages}{1867--1876}.
\newblock
\showISBNx{978-1-4503-4655-9}
\urldef\tempurl%
\url{https://doi.org/10.1145/3025453.3025687}
\showDOI{\tempurl}


\bibitem[Nie et~al\mbox{.}(2020)]%
        {nie_chaos_2020}
\bibfield{author}{\bibinfo{person}{Yixin Nie}, \bibinfo{person}{Xiang Zhou},
  {and} \bibinfo{person}{Mohit Bansal}.} \bibinfo{year}{2020}\natexlab{}.
\newblock \showarticletitle{What {Can} {We} {Learn} from {Collective} {Human}
  {Opinions} on {Natural} {Language} {Inference} {Data}?}
\newblock \bibinfo{journal}{\emph{arXiv:2010.03532 [cs]}} (\bibinfo{date}{Oct.}
  \bibinfo{year}{2020}).
\newblock
\urldef\tempurl%
\url{http://arxiv.org/abs/2010.03532}
\showURL{%
\tempurl}
\newblock
\shownote{arXiv: 2010.03532}.


\bibitem[Ohn and Ohn(2019)]%
        {ohn_evaluation_2019}
\bibfield{author}{\bibinfo{person}{May~Honey Ohn} {and}
  \bibinfo{person}{Khin-Maung Ohn}.} \bibinfo{year}{2019}\natexlab{}.
\newblock \showarticletitle{An evaluation study on gamified online learning
  experiences and its acceptance among medical students}.
\newblock \bibinfo{journal}{\emph{Tzu-Chi Medical Journal}}
  \bibinfo{volume}{32}, \bibinfo{number}{2} (\bibinfo{date}{June}
  \bibinfo{year}{2019}), \bibinfo{pages}{211--215}.
\newblock
\showISSN{1016-3190}
\urldef\tempurl%
\url{https://doi.org/10.4103/tcmj.tcmj_5_19}
\showDOI{\tempurl}


\bibitem[Oviatt(2006)]%
        {oviatt_human-centered_2006}
\bibfield{author}{\bibinfo{person}{Sharon Oviatt}.}
  \bibinfo{year}{2006}\natexlab{}.
\newblock \showarticletitle{Human-centered design meets cognitive load theory:
  designing interfaces that help people think}. In
  \bibinfo{booktitle}{\emph{Proceedings of the 14th {ACM} international
  conference on {Multimedia}}} \emph{(\bibinfo{series}{{MM} '06})}.
  \bibinfo{publisher}{Association for Computing Machinery},
  \bibinfo{address}{New York, NY, USA}, \bibinfo{pages}{871--880}.
\newblock
\showISBNx{978-1-59593-447-5}
\urldef\tempurl%
\url{https://doi.org/10.1145/1180639.1180831}
\showDOI{\tempurl}


\bibitem[Rajpurkar et~al\mbox{.}(2016)]%
        {rajpurkar-etal-2016-squad}
\bibfield{author}{\bibinfo{person}{Pranav Rajpurkar}, \bibinfo{person}{Jian
  Zhang}, \bibinfo{person}{Konstantin Lopyrev}, {and} \bibinfo{person}{Percy
  Liang}.} \bibinfo{year}{2016}\natexlab{}.
\newblock \showarticletitle{{SQ}u{AD}: 100,000+ Questions for Machine
  Comprehension of Text}. In \bibinfo{booktitle}{\emph{Proceedings of the 2016
  Conference on Empirical Methods in Natural Language Processing}}.
  \bibinfo{publisher}{Association for Computational Linguistics},
  \bibinfo{address}{Austin, Texas}, \bibinfo{pages}{2383--2392}.
\newblock
\urldef\tempurl%
\url{https://doi.org/10.18653/v1/D16-1264}
\showDOI{\tempurl}


\bibitem[Salminen et~al\mbox{.}(2018)]%
        {salminen_anatomy_nodate}
\bibfield{author}{\bibinfo{person}{Joni~O. Salminen}, \bibinfo{person}{Hind
  Almerekhi}, \bibinfo{person}{Milica Milenkovic}, \bibinfo{person}{Soon-Gyo
  Jung}, \bibinfo{person}{Jisun An}, \bibinfo{person}{Haewoon Kwak}, {and}
  \bibinfo{person}{Bernard~Jim Jansen}.} \bibinfo{year}{2018}\natexlab{}.
\newblock \showarticletitle{Anatomy of Online Hate: Developing a Taxonomy and
  Machine Learning Models for Identifying and Classifying Hate in Online News
  Media}. In \bibinfo{booktitle}{\emph{International Conference on Web and
  Social Media}}.
\newblock


\bibitem[Seong and Kim(2020)]%
        {seong_designing_2020}
\bibfield{author}{\bibinfo{person}{Eunjin Seong} {and}
  \bibinfo{person}{Seungjun Kim}.} \bibinfo{year}{2020}\natexlab{}.
\newblock \showarticletitle{Designing a {Crowdsourcing} {System} for the
  {Elderly}: {A} {Gamified} {Approach} to {Speech} {Collection}}. In
  \bibinfo{booktitle}{\emph{Extended {Abstracts} of the 2020 {CHI} {Conference}
  on {Human} {Factors} in {Computing} {Systems}}} \emph{(\bibinfo{series}{{CHI}
  {EA} '20})}. \bibinfo{publisher}{Association for Computing Machinery},
  \bibinfo{address}{New York, NY, USA}, \bibinfo{pages}{1--9}.
\newblock
\showISBNx{978-1-4503-6819-3}
\urldef\tempurl%
\url{https://doi.org/10.1145/3334480.3382999}
\showDOI{\tempurl}


\bibitem[Shah et~al\mbox{.}(2021)]%
        {shah_permutation-based_2021}
\bibfield{author}{\bibinfo{person}{Nihar~B. Shah}, \bibinfo{person}{Sivaraman
  Balakrishnan}, {and} \bibinfo{person}{Martin~J. Wainwright}.}
  \bibinfo{year}{2021}\natexlab{}.
\newblock \showarticletitle{A {Permutation}-based {Model} for {Crowd}
  {Labeling}: {Optimal} {Estimation} and {Robustness}}.
\newblock \bibinfo{journal}{\emph{IEEE Transactions on Information Theory}}
  \bibinfo{volume}{67}, \bibinfo{number}{6} (\bibinfo{date}{June}
  \bibinfo{year}{2021}), \bibinfo{pages}{4162--4184}.
\newblock
\showISSN{0018-9448, 1557-9654}
\urldef\tempurl%
\url{https://doi.org/10.1109/TIT.2020.3045613}
\showDOI{\tempurl}
\newblock
\shownote{arXiv:1606.09632 [cs, math, stat]}.


\bibitem[Sigurdsson et~al\mbox{.}(2016)]%
        {sigurdsson_much_2016}
\bibfield{author}{\bibinfo{person}{Gunnar~A. Sigurdsson}, \bibinfo{person}{Olga
  Russakovsky}, \bibinfo{person}{Ali Farhadi}, \bibinfo{person}{Ivan Laptev},
  {and} \bibinfo{person}{Abhinav Gupta}.} \bibinfo{year}{2016}\natexlab{}.
\newblock \bibinfo{title}{Much {Ado} {About} {Time}: {Exhaustive} {Annotation}
  of {Temporal} {Data}}.
\newblock
\newblock
\urldef\tempurl%
\url{https://doi.org/10.48550/arXiv.1607.07429}
\showDOI{\tempurl}
\newblock
\shownote{arXiv:1607.07429 [cs]}.


\bibitem[Stureborg et~al\mbox{.}(2023)]%
        {vax_taxonomy}
\bibfield{author}{\bibinfo{person}{Rickard Stureborg}, \bibinfo{person}{Bhuwan
  Dhingra}, \bibinfo{person}{Jun Yang}, {and} \bibinfo{person}{Lavanya
  Vasudevan}.} \bibinfo{year}{2023}\natexlab{}.
\newblock \bibinfo{title}{Development and validation of VaxConcerns: a taxonomy
  for vaccine concerns with crowdsource-viability}.
\newblock
\newblock
\urldef\tempurl%
\url{http://rickard.stureborg.com/papers/vax_taxonomy}
\showURL{%
\tempurl}


\bibitem[Swanson et~al\mbox{.}(2015)]%
        {swanson_snapshot_2015}
\bibfield{author}{\bibinfo{person}{Alexandra Swanson},
  \bibinfo{person}{Margaret Kosmala}, \bibinfo{person}{Chris Lintott},
  \bibinfo{person}{Robert Simpson}, \bibinfo{person}{Arfon Smith}, {and}
  \bibinfo{person}{Craig Packer}.} \bibinfo{year}{2015}\natexlab{}.
\newblock \showarticletitle{Snapshot {Serengeti}, high-frequency annotated
  camera trap images of 40 mammalian species in an {African} savanna}.
\newblock \bibinfo{journal}{\emph{Scientific Data}} \bibinfo{volume}{2},
  \bibinfo{number}{1} (\bibinfo{date}{June} \bibinfo{year}{2015}),
  \bibinfo{pages}{150026}.
\newblock
\showISSN{2052-4463}
\urldef\tempurl%
\url{https://doi.org/10.1038/sdata.2015.26}
\showDOI{\tempurl}
\newblock
\shownote{Number: 1 Publisher: Nature Publishing Group}.


\bibitem[Sweller(2011)]%
        {sweller_chapter_2011}
\bibfield{author}{\bibinfo{person}{John Sweller}.}
  \bibinfo{year}{2011}\natexlab{}.
\newblock \showarticletitle{{CHAPTER} {TWO} - {Cognitive} {Load} {Theory}}.
\newblock In \bibinfo{booktitle}{\emph{Psychology of {Learning} and
  {Motivation}}}, \bibfield{editor}{\bibinfo{person}{Jose~P. Mestre} {and}
  \bibinfo{person}{Brian~H. Ross}} (Eds.). Vol.~\bibinfo{volume}{55}.
  \bibinfo{publisher}{Academic Press}, \bibinfo{pages}{37--76}.
\newblock
\urldef\tempurl%
\url{https://doi.org/10.1016/B978-0-12-387691-1.00002-8}
\showDOI{\tempurl}


\bibitem[Thorne et~al\mbox{.}(2018)]%
        {thorne_fever_2018}
\bibfield{author}{\bibinfo{person}{James Thorne}, \bibinfo{person}{Andreas
  Vlachos}, \bibinfo{person}{Christos Christodoulopoulos}, {and}
  \bibinfo{person}{Arpit Mittal}.} \bibinfo{year}{2018}\natexlab{}.
\newblock \showarticletitle{{FEVER}: a large-scale dataset for {Fact}
  {Extraction} and {VERification}}.
\newblock \bibinfo{journal}{\emph{arXiv:1803.05355 [cs]}} (\bibinfo{date}{Dec.}
  \bibinfo{year}{2018}).
\newblock
\urldef\tempurl%
\url{http://arxiv.org/abs/1803.05355}
\showURL{%
\tempurl}
\newblock
\shownote{arXiv: 1803.05355}.


\bibitem[Van~Horn et~al\mbox{.}(2018)]%
        {inaturalist}
\bibfield{author}{\bibinfo{person}{Grant Van~Horn}, \bibinfo{person}{Oisin
  Mac~Aodha}, \bibinfo{person}{Yang Song}, \bibinfo{person}{Yin Cui},
  \bibinfo{person}{Chen Sun}, \bibinfo{person}{Alex Shepard},
  \bibinfo{person}{Hartwig Adam}, \bibinfo{person}{Pietro Perona}, {and}
  \bibinfo{person}{Serge Belongie}.} \bibinfo{year}{2018}\natexlab{}.
\newblock \showarticletitle{The inaturalist species classification and
  detection dataset}. In \bibinfo{booktitle}{\emph{Proceedings of the IEEE
  conference on computer vision and pattern recognition}}.
  \bibinfo{pages}{8769--8778}.
\newblock


\bibitem[Vondrick et~al\mbox{.}(2013)]%
        {vondrick_efficiently_2013}
\bibfield{author}{\bibinfo{person}{Carl Vondrick}, \bibinfo{person}{Donald
  Patterson}, {and} \bibinfo{person}{Deva Ramanan}.}
  \bibinfo{year}{2013}\natexlab{}.
\newblock \showarticletitle{Efficiently {Scaling} up {Crowdsourced} {Video}
  {Annotation}}.
\newblock \bibinfo{journal}{\emph{International Journal of Computer Vision}}
  \bibinfo{volume}{101}, \bibinfo{number}{1} (\bibinfo{date}{Jan.}
  \bibinfo{year}{2013}), \bibinfo{pages}{184--204}.
\newblock
\showISSN{0920-5691}
\urldef\tempurl%
\url{https://doi.org/10.1007/s11263-012-0564-1}
\showDOI{\tempurl}


\bibitem[Whitehill et~al\mbox{.}(2009)]%
        {whitehill_whose_2009}
\bibfield{author}{\bibinfo{person}{Jacob Whitehill}, \bibinfo{person}{Ting-fan
  Wu}, \bibinfo{person}{Jacob Bergsma}, \bibinfo{person}{Javier Movellan},
  {and} \bibinfo{person}{Paul Ruvolo}.} \bibinfo{year}{2009}\natexlab{}.
\newblock \showarticletitle{Whose {Vote} {Should} {Count} {More}: {Optimal}
  {Integration} of {Labels} from {Labelers} of {Unknown} {Expertise}}. In
  \bibinfo{booktitle}{\emph{Advances in {Neural} {Information} {Processing}
  {Systems}}}, Vol.~\bibinfo{volume}{22}. \bibinfo{publisher}{Curran
  Associates, Inc.}
\newblock
\urldef\tempurl%
\url{https://papers.nips.cc/paper/2009/hash/f899139df5e1059396431415e770c6dd-Abstract.html}
\showURL{%
\tempurl}


\bibitem[Wickelgren(1977)]%
        {wickelgren_speed-accuracy_1977}
\bibfield{author}{\bibinfo{person}{Wayne~A. Wickelgren}.}
  \bibinfo{year}{1977}\natexlab{}.
\newblock \showarticletitle{Speed-accuracy tradeoff and information processing
  dynamics}.
\newblock \bibinfo{journal}{\emph{Acta Psychologica}} \bibinfo{volume}{41},
  \bibinfo{number}{1} (\bibinfo{date}{Feb.} \bibinfo{year}{1977}),
  \bibinfo{pages}{67--85}.
\newblock
\showISSN{0001-6918}
\urldef\tempurl%
\url{https://doi.org/10.1016/0001-6918(77)90012-9}
\showDOI{\tempurl}


\bibitem[Williams et~al\mbox{.}(2018)]%
        {multinli}
\bibfield{author}{\bibinfo{person}{Adina Williams}, \bibinfo{person}{Nikita
  Nangia}, {and} \bibinfo{person}{Samuel Bowman}.}
  \bibinfo{year}{2018}\natexlab{}.
\newblock \showarticletitle{A Broad-Coverage Challenge Corpus for Sentence
  Understanding through Inference}. In \bibinfo{booktitle}{\emph{Proceedings of
  the 2018 Conference of the North American Chapter of the Association for
  Computational Linguistics: Human Language Technologies, Volume 1 (Long
  Papers)}} (New Orleans, Louisiana). \bibinfo{publisher}{Association for
  Computational Linguistics}, \bibinfo{pages}{1112--1122}.
\newblock
\urldef\tempurl%
\url{http://aclweb.org/anthology/N18-1101}
\showURL{%
\tempurl}


\bibitem[Yu et~al\mbox{.}(2016)]%
        {yu_lsun_2016}
\bibfield{author}{\bibinfo{person}{Fisher Yu}, \bibinfo{person}{Ari Seff},
  \bibinfo{person}{Yinda Zhang}, \bibinfo{person}{Shuran Song},
  \bibinfo{person}{Thomas Funkhouser}, {and} \bibinfo{person}{Jianxiong Xiao}.}
  \bibinfo{year}{2016}\natexlab{}.
\newblock \bibinfo{title}{{LSUN}: {Construction} of a {Large}-scale {Image}
  {Dataset} using {Deep} {Learning} with {Humans} in the {Loop}}.
\newblock
\newblock
\urldef\tempurl%
\url{https://doi.org/10.48550/arXiv.1506.03365}
\showDOI{\tempurl}
\newblock
\shownote{arXiv:1506.03365 [cs] version: 3}.


\bibitem[Zhang et~al\mbox{.}(2018)]%
        {zhang_ontological_2018}
\bibfield{author}{\bibinfo{person}{Jingpu Zhang}, \bibinfo{person}{Zuping
  Zhang}, \bibinfo{person}{Zixiang Wang}, \bibinfo{person}{Yuting Liu}, {and}
  \bibinfo{person}{Lei Deng}.} \bibinfo{year}{2018}\natexlab{}.
\newblock \showarticletitle{Ontological function annotation of long non-coding
  {RNAs} through hierarchical multi-label classification}.
\newblock \bibinfo{journal}{\emph{Bioinformatics}} \bibinfo{volume}{34},
  \bibinfo{number}{10} (\bibinfo{date}{May} \bibinfo{year}{2018}),
  \bibinfo{pages}{1750--1757}.
\newblock
\showISSN{1367-4803}
\urldef\tempurl%
\url{https://doi.org/10.1093/bioinformatics/btx833}
\showDOI{\tempurl}


\bibitem[Zhang et~al\mbox{.}(2019)]%
        {zhang_text_2019}
\bibfield{author}{\bibinfo{person}{Mingrui~Ray Zhang}, \bibinfo{person}{Shumin
  Zhai}, {and} \bibinfo{person}{Jacob~O. Wobbrock}.}
  \bibinfo{year}{2019}\natexlab{}.
\newblock \showarticletitle{Text {Entry} {Throughput}: {Towards} {Unifying}
  {Speed} and {Accuracy} in a {Single} {Performance} {Metric}}. In
  \bibinfo{booktitle}{\emph{Proceedings of the 2019 {CHI} {Conference} on
  {Human} {Factors} in {Computing} {Systems}}} \emph{(\bibinfo{series}{{CHI}
  '19})}. \bibinfo{publisher}{Association for Computing Machinery},
  \bibinfo{address}{New York, NY, USA}, \bibinfo{pages}{1--13}.
\newblock
\showISBNx{978-1-4503-5970-2}
\urldef\tempurl%
\url{https://doi.org/10.1145/3290605.3300866}
\showDOI{\tempurl}


\bibitem[Zhou et~al\mbox{.}(2021)]%
        {zhou_distributed_2021}
\bibfield{author}{\bibinfo{person}{Xiang Zhou}, \bibinfo{person}{Yixin Nie},
  {and} \bibinfo{person}{Mohit Bansal}.} \bibinfo{year}{2021}\natexlab{}.
\newblock \showarticletitle{Distributed {NLI}: {Learning} to {Predict} {Human}
  {Opinion} {Distributions} for {Language} {Reasoning}}.
\newblock \bibinfo{journal}{\emph{arXiv:2104.08676 [cs]}}
  (\bibinfo{date}{April} \bibinfo{year}{2021}).
\newblock
\urldef\tempurl%
\url{http://arxiv.org/abs/2104.08676}
\showURL{%
\tempurl}
\newblock
\shownote{arXiv: 2104.08676}.


\end{thebibliography}

\appendix

\section{The Vaccine Concerns (VaxConcerns) Taxonomy}
\label{apx:taxonomy}
\begin{figure}[H]
   \centering
   \includegraphics[width=.40\textwidth]{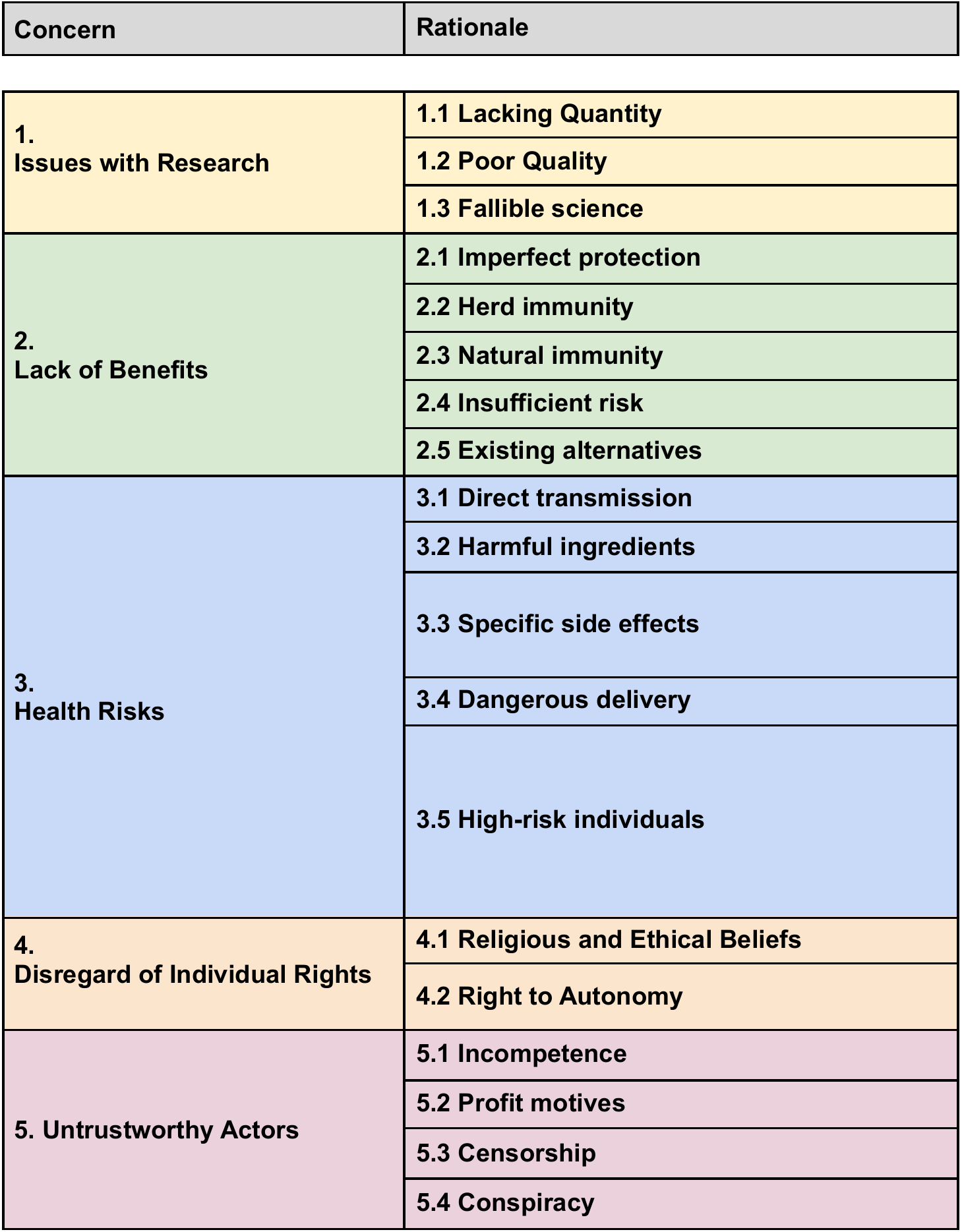}
 \caption{``VaxConcerns'' taxonomy used to label all passages in the experiments}
\Description{A bar chart is shown giving the value for worker F1 score on the X axis ranging from 0 to 1, and the difficulty on the Y axis. The labels, from top to bottom, on the Y axis read 'immediate author agreement', 'author agreement after providing rationale', and 'author agreement after discussion'.}
\end{figure}

\section{Alternative Text for Figures}
\textbf{Figure 1:} “Three diagrams are shown describing single-pass, multi-pass, and hierarchical multi-pass routing logic. For single-pass, all set of labels are given to one worker. For multi-pass, the labels are partitioned into groups (1,2,...) and given to separate workers (A,B,...). For hierarchical multi-label, the top level labels are given to one worker, who's annotations determine whether or not the child labels will be given as a new group to annotators downstream. This example shows the case where the first worker labels TFFTF, and downstream there are two tasks set up for new workers to label the children of label 1 and label 4, respectively.”
\textbf{Figure 2:} “The figure shows an example passage that reads: 'The minister of fear (the CDC) was working overtime peddling doom and gloom, knowing that frightened people do not make rational decisions --- nothing sells vaccines like panic.'”
\textbf{Figure 3:} “Four diagrams are shown side by side. In each diagram there are a set of checkboxes or radio buttons indicating how the labels will be presented to the user. Binary label (the leftmost diagram) contains a simple question '1.1?' and below it a radio button reading 'yes' or 'no'. Multi-label contains a simple list of checkboxes labeled '1.1, 1.2, ...'. Hierarchical multi-label v1 contains staggered checkboxes where the leftmost checkboxes read '1, 2' and the boxes immediately under these are tucked under them, reading '1.1, ...' for the parent label '1', and '2.1, ...' for the parent label '2'. Hierarchical multi-label v2 contains both the radio button setup from the leftmost diagram, as well as the checkboxes from multi-label underneath them.”
\textbf{Figure 4:} “A table is shown with column headers reading 'interface design', 'greater than or equal to 1 tutorial Q', 'greater than or equal to full tutorial', 'greater than or equal to took exam', and 'greater than or equal to 1 datapoint'. The table shows values for all 6 labeling schemes.”
\textbf{Figure 5:} “A table is shown with values for precision, recall, and F1 score. These metrics are given for each of the 6 labeling schemes, and a random baseline is shown at the bottom. Hrchl-pass multi has bold font at the f1 score indicating it is the highest value in that column: 0.56.”
\textbf{Figure 6:} “A two-part table is shown with column headers 'model factor', 'estimate', '95$\%$ CI', 'SE', and 'p-value'. The first part of the table (top) has a subheading that reads 'labeling\_scheme (baseline=multi-pass random multi-label)', and the second part of the table (bottom) has a subheading that reads 'additional numerical factors'. The first part includes 5 of the 6 labeling schemes, while the bottom includes new factors such as 'time\_started', 'percentage\_easy', and 'true\_positive\_freq'. Some values in the table are denoted '*' which represents a p-value below 0.001.”
\textbf{Figure 7:} “The figure shows an example passage that reads: 'Pregnant women given vaccine have babies with more health problems'”
\textbf{Figure 8:} “A bar chart is shown giving the value for worker F1 score on the X axis ranging from 0 to 1, and the difficulty on the Y axis. The labels, from top to bottom, on the Y axis read 'immediate author agreement', 'author agreement after providing rationale', and 'author agreement after discussion'.”
\textbf{Figure 9:} “A scatter plot shows orange and blue dots generally following a linearly positive relationship. The orange dots are labeled 'hrchl-pass' and the blue dots 'multi-pass'. On the X axis: 'Frequency of True Positives shown to workers'. On the Y axis: Worker F1. Each blue dot is paired with an orange dot through an arrow which is drawn between them pointing towards the orange dot.”
\textbf{Figure 10:} “A line plot is shown with dashed lines between three dots. The dots are lined up at tickmarks labeled 'sensitive', 'majority', and 'unanimous'. This is repeated for all 6 labeling schemes, leaving 6 connected dotted lines all in different colors. Every one of the lines follows an inverted V shape, with their highest point being over the 'majority' tick mark.”

\section{Degradation in worker task performance over time}
\label{apx:degradation}
\begin{figure}[H]
   \centering
   \includegraphics[width=.45\textwidth]{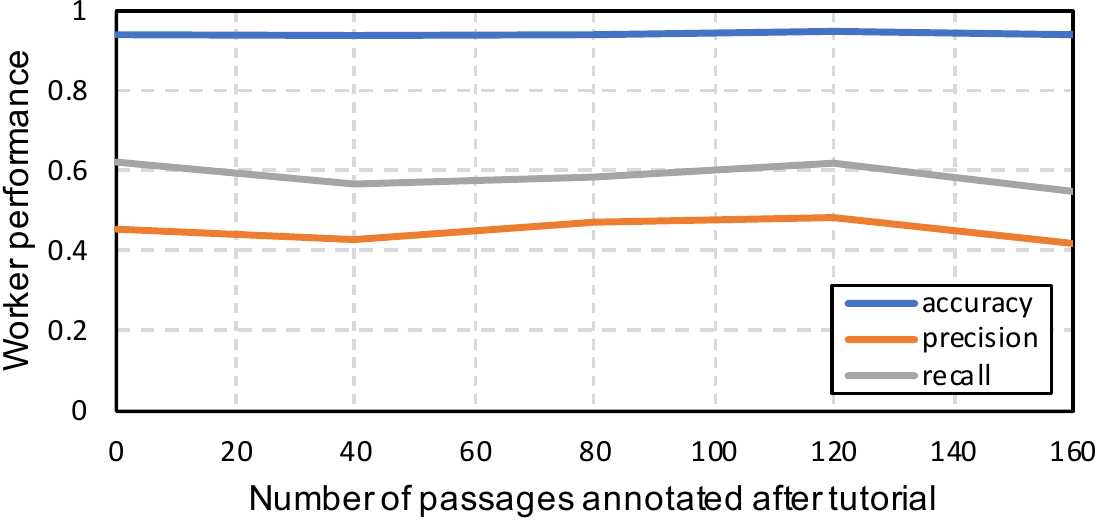}
 \caption{
    Mean performance of workers as they annotate passages.
    Worker performance fluctuates very little as they gain experience on the platform.}
\Description{A scatter plot shows orange and blue dots generally following a linearly positive relationship. The orange dots are labeled 'hrchl-pass' and the blue dots 'multi-pass'. On the X axis: 'Frequency of True Positives shown to workers'. On the Y axis: Worker F1. Each blue dot is paired with an orange dot through an arrow which is drawn between them pointing towards the orange dot.}
 \label{fig:degradation}
\end{figure}

\section{Distributions of time spent labeling each passage}
\label{apx:time_distributions}
\begin{figure}[H]
   \centering
   \includegraphics[width=.45\textwidth]{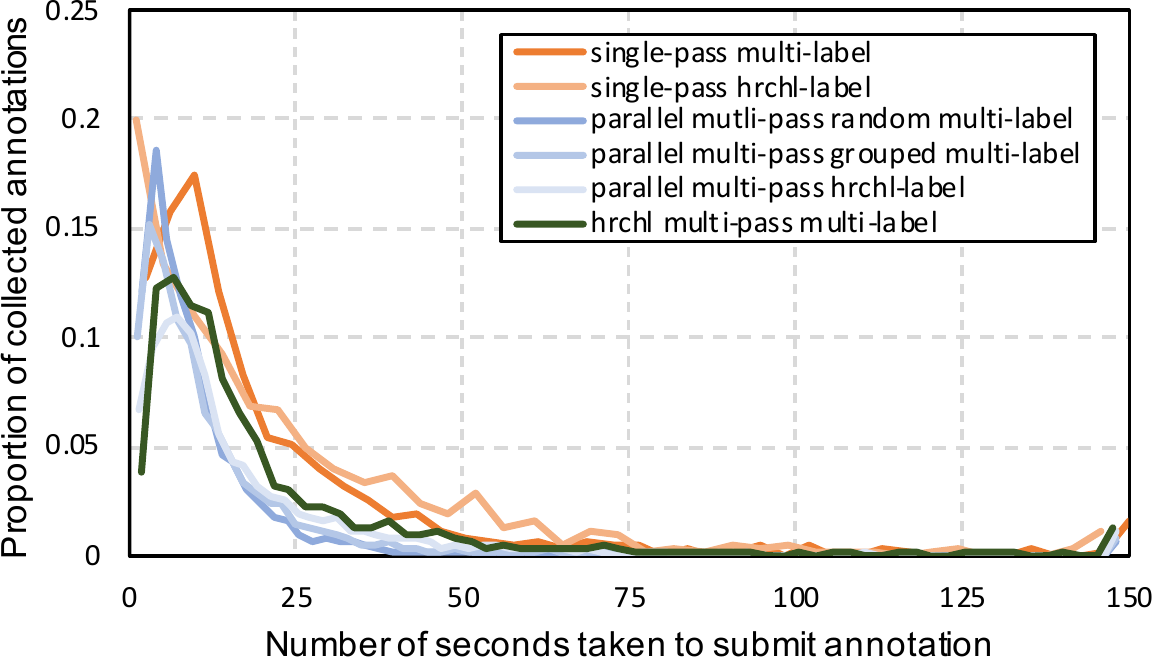}
 \caption{Distributions for amount of time spent labeling each passage, broken down by each interface design. Values are limited at a maximum of 2.5 minutes to account for behavior such as stepping away from the task to take a break.}
\Description{A line plot is shown with dashed lines between three dots. The dots are lined up at tickmarks labeled 'sensitive', 'majority', and 'unanimous'. This is repeated for all 6 labeling schemes, leaving 6 connected dotted lines all in different colors. Every one of the lines follows an inverted V shape, with their highest point being over the 'majority' tick mark.}
 \label{fig:time_spent}
\end{figure}

\section{Screenshot of Definitions Task in Annotation Platform}
\label{apx:definitions_task}
\begin{figure}[H]
   \centering
   \includegraphics[width=.45\textwidth]{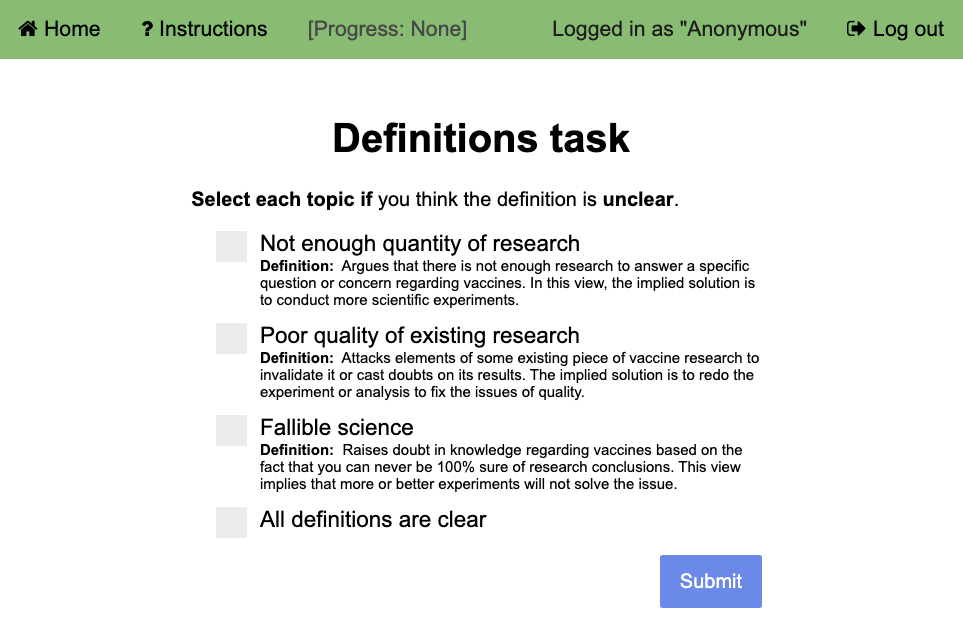}
 \caption{Definitions task presented to workers before they annotate any passages. \new{Note that we are using a custom annotation platform to achieve a higher level control than offered by AMT with respect to quality checks, batching multiple passages for a single hit, and measuring worker behaviors, among other things.}}
\end{figure}

\section{Methods for Pre-Processing Data}
\new{We ignore all data collected for tutorial paragraphs, entrance exam, and quality checks when assessing workers. In the future we may assess how to leverage any information about worker performance on the tutorial towards improving data collection quality.}

\new{We remove all data collected from a worker if they failed the entrance exam or a quality check. Such workers were also banned in live time during data collection to avoid spending any further of our research budget on their data collection. To be clear, when setting the budgets for each labeling scheme, we do not factor in additional cost due to such workers. Rather, we re-annotate all the passages which they had been paid for.}

\label{label_consistency}
\new{
For certain labeling schemes, consistency between level 1 and level 2 is forced by the annotation platform. For example, in single-pass hrchl-label, workers cannot submit a positive value for a level 2 label without also submitting a positive value for its parent node in the taxonomy. This is achieved through some javascript in the annotation platform and is visually confirmed to the workers while labeling. However, for other labeling schemes such as multi-pass multi-label, the separation of L1 and L2 labels into separate screens leads to collecting data which is not necessarily consistent. Since any real-world use of this latter type of labeling scheme would include corrections for consistency, we correct for this during post-processing. This ensures fair comparison between the labeling schemes such that we don’t disadvantage multi-pass schemes. To further ensure comparison is consistent, we make sure to look at performance on level 2 labels on its own during our analysis.
}

\section{Metrics}
\label{apx:metrics}
\new{
We measure annotator performance using the F1 score, a harmonic mean of precision and recall. This score is commonly used in machine learning tasks and is a well understood and cared about metric in research communities such as natural language processing (NLP).
Examining this metric gives more utility for NLP researchers, since the positively labeled examples will be the most informative during model training. One can achieve very high accuracy simply by marking all passages as negative.
Since each worker’s performance can be evaluated across each label in the taxonomy individually, we must use some aggregation technique when presenting our scores. 
We employ a macro-level average of F1, which is computed by first finding the F1 score on every taxonomy label, and then averaging across all these labels.
In the case where we give an F1 score for each worker, the macro-averaging process happens in parallel for each worker. 
In the case where we give a single F1 score for all workers, we take all the annotations and treat them as if a single worker had filled out all annotations and we then follow the macro-averaging process.
This method of averaging is preferable in our setting as opposed to a samples-based average which would compute F1 over a single passage, and then aggregate across passages. Semantically, we use this method since we also care about worker performance on each individual label and can thereby inspect these values. Inspecting worker performance on each passage is less important to us, since the set of passages used are meant to be a sample of the types of passages annotated in any application of our work.
}

\section{Task uptake}
\label{sec:uptake}
We train each worker before they complete any real annotations.
Workers get paid for this training in order to ensure a fair and non-predatory (\cite{barbosa_rehumanized_2019}) employment.
This leads to additional costs to those interested in paying for the data labeling, since some workers can go through some or even all of the training, yet submit no actual annotations.
Such workers never become ``productive.''
Below we report the task uptake, that is the percentage of workers who became productive, out of all the workers which completed at least one tutorial example.
We also report an inefficiency number, which is the percentage of extraneous annotations collected from training and attention checks (discussed in detail in \S\ref{sec:training}).
The inefficiency number compares the extraneous annotations to the total number of ``useful'' annotations collected.
For multi-pass schemes, the workers were allowed to complete one partition (group) of the labels and then begin a new partition. 
This occurred seamlessly, whereas there was a multi-day delay for the \textit{hrchl-pass multi} scheme, meaning that there were less return workers for this task and thus decreasing the total task uptake.\footnote{This multi-day delay was due to the annotation platform we used not supporting this immediate switch when the first labeling scheme was deployed.}

Task inefficiency (which is directly proportional to additional incurred cost) is lowest for \textit{hrchl-pass multi}.
The task uptake is the highest for \textit{multi-pass random multi} at $80\%$, which contrasts the performance trends shown in \S\ref{sec:performance}.
One potential explanation for this could be that annotators underestimate the difficulty of the task when they're presented randomly partitioned labels.

\begin{table}[H]
\begin{tabular}{@{}c@{}}
   \includegraphics[width=.35\textwidth]{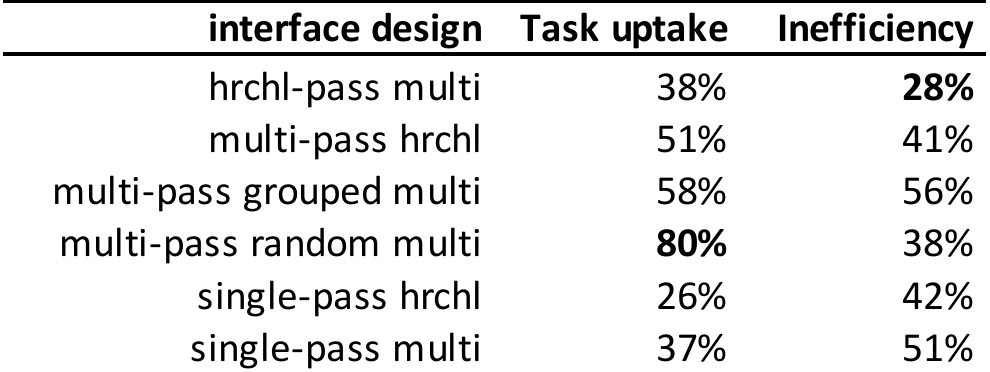}
\end{tabular}
 \caption{
    Data collection inefficiencies due to training examples completed by workers. 
    Uptake shows the percentage of workers who complete at least one ``real'' passage, 
    while inefficiency shows the total percentage of extraneous annotations collected.
    Inefficiency is the lowest using a hrchl-pass scheme}
 \label{fig:uptake}
\end{table}

\section{True positive frequency vs F1 scores}
\label{apx:tp_freq_all}
\begin{figure}[H]
   \centering
   \includegraphics[width=.45\textwidth]{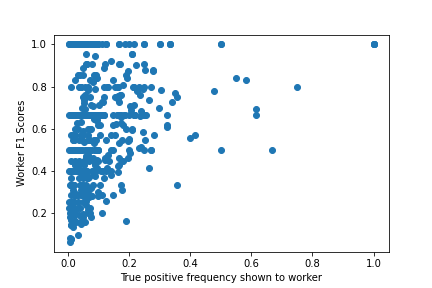}
 \caption{
    \new{
    Relationship between the true positive frequency of each worker and their F1 scores.
        }}
\end{figure}

\section{Worker Agreement on Labels by Labeling Scheme}
\label{apx:agreement}
\begin{table}[H]
  \begin{tabular}{@{}c@{}}
   \includegraphics[width=.45\textwidth]{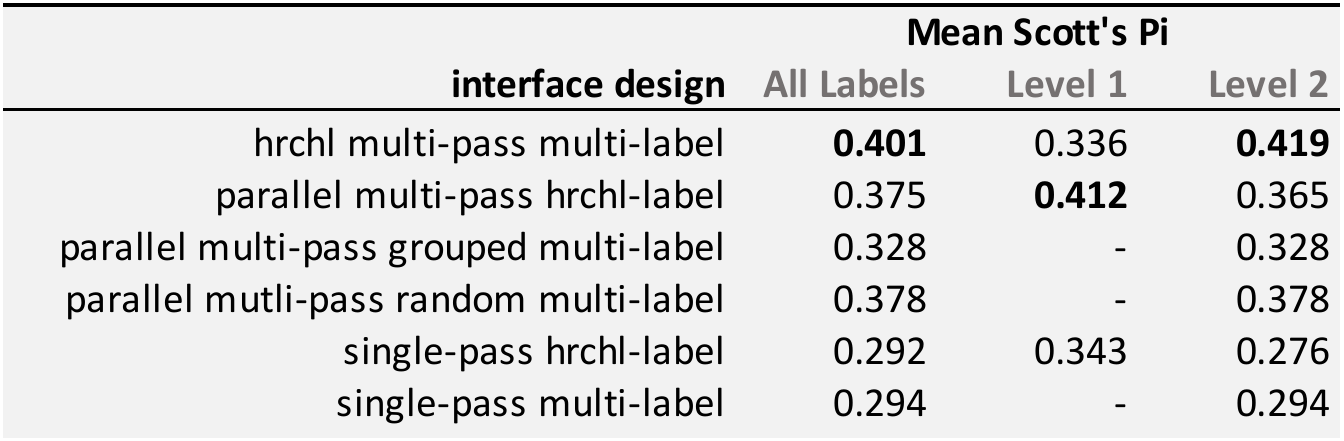}
  \end{tabular}
 \caption{Mean Scott's Pi label agreement between crowdsource workers in each labeling scheme}
 \label{fig:agreement}
\end{table}

\section{Bootstrap confidence intervals}
\label{apx:conf_intervals}
\new{To find the confidence intervals included in any results of the paper, we use bootstrap confidence intervals. This is done directly on the raw data we collected, where a datapoint is a single submission by a worker. That is, for single-pass schemes the datapoint will include all labels from the taxonomy, but for multi-pass schemes the datapoint will only include a subset of the labels. We draw N=10,000 samples with replacement from the original data, then find the performance metric using the process outlined in the paper (including all preprocessing). We then use these 10,000 measurements of each performance metric to compute 99\% confidence intervals.}

\new{We do this sampling on the subset of annotations which ultimately contribute to the performance metric, rather than all the annotations we receive. This ensures we are not sampling (for example) tutorial annotations which we will ignore in the final calculations anyway.}
\begin{figure}[H]
   \centering
   \includegraphics[width=0.45\textwidth]{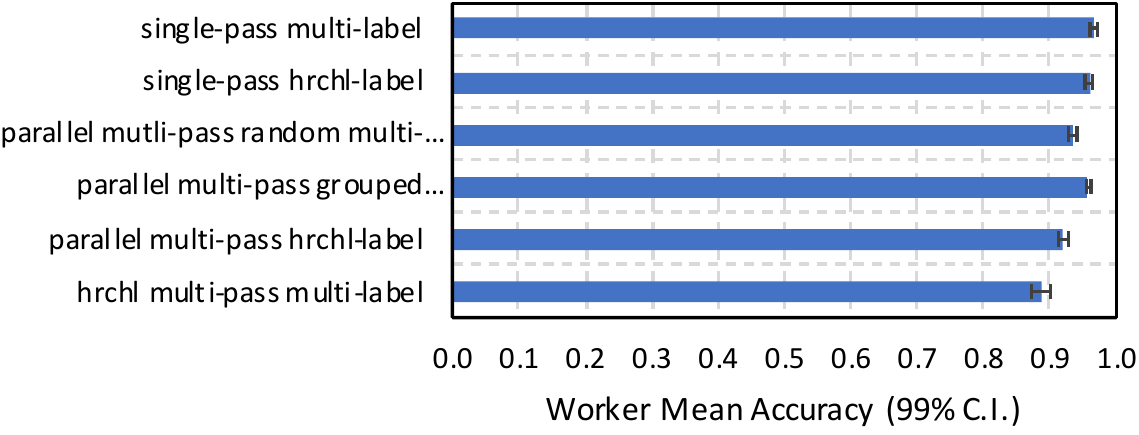} \\
   \vspace{0.2cm}
   \includegraphics[width=0.45\textwidth]{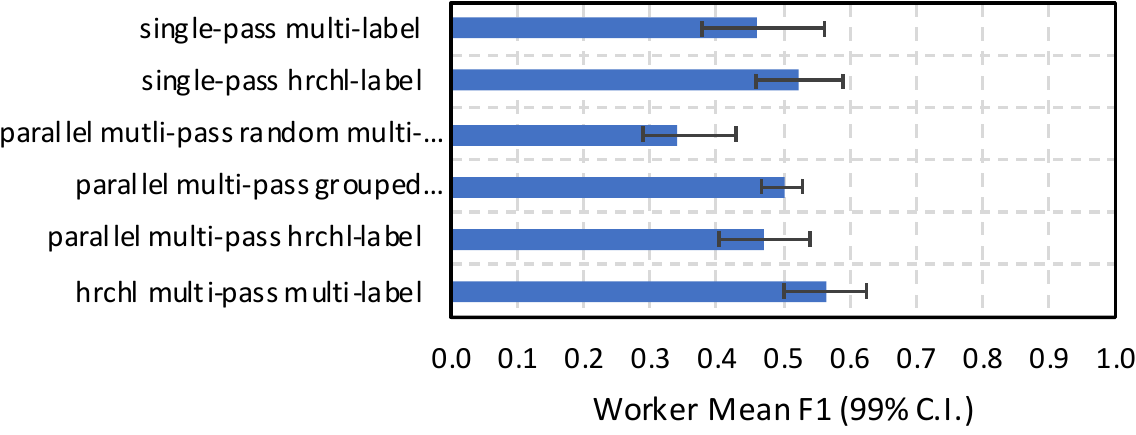} \\
   \vspace{0.2cm}
   \includegraphics[width=0.45\textwidth]{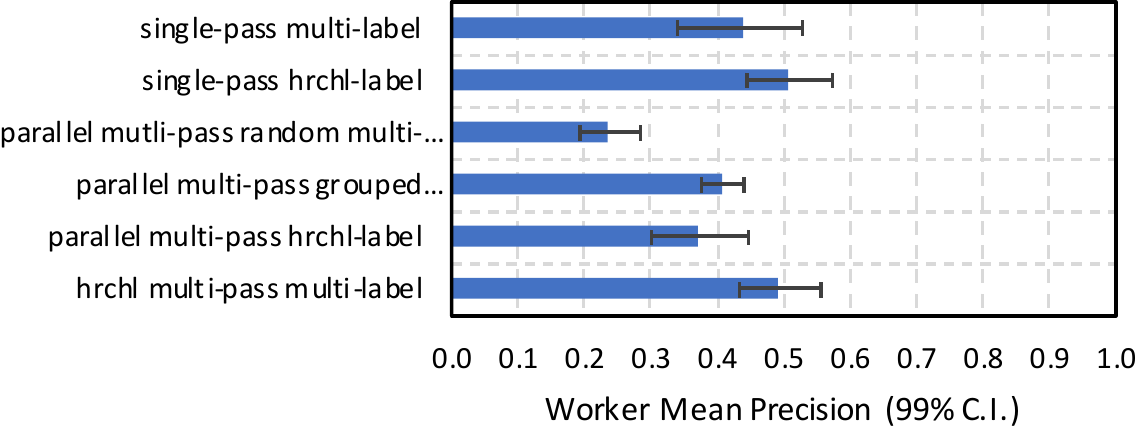} \\
   \vspace{0.2cm}
   \includegraphics[width=0.45\textwidth]{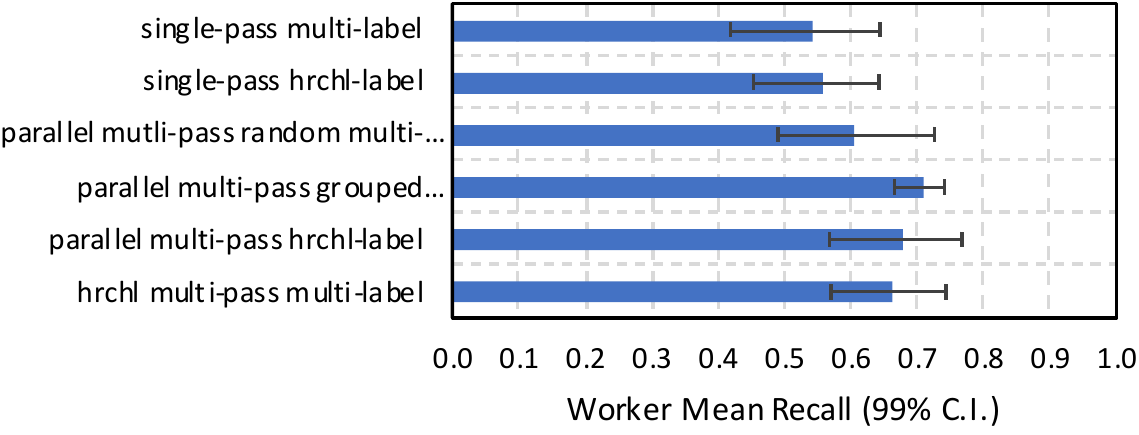}
 \caption{\new{Confidence intervals for worker mean performance metrics.}}
\end{figure}

\section{Toy Example of Annotation Costs}
\label{apx:cost}
Suppose the team wants to collect approximately $10{,}000$ fully labeled passages in order to provide high-quality training data.
How much will it cost them to use each labeling scheme?

\begin{figure}[H]
    \centering
    \includegraphics[width=.45\textwidth]{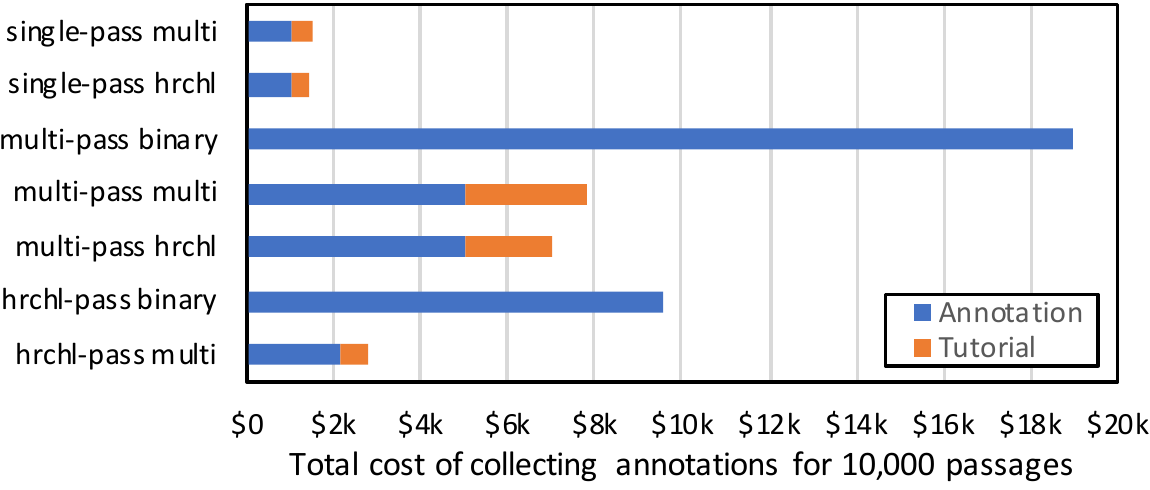}
    \caption{
    Cost for labeling 10,000 passages under various labeling schemes, assuming a reward of $\$0.10$ per passage. 
    Orange bars show the cost of training workers (further details in §\ref{sec:uptake}).
    \new{
    Note that we do not show orange bars for the binary-label setups since we were not able to run these experiments and estimate the average overhead of training workers. Despite this, binary labeling is prohibitively expensive: 8.8x the cost of a hrchl-pass multi-label scheme.
    }}
    \label{fig:cost}
\end{figure}

\new{
If the team assumes the longest part of labeling is due to reading, and want to guarantee a strong hourly wage for workers, then they can fix the reward at \$0.10 per passage (this is a toy example, but loosely this is in line with paying above minimum wage in the United States).
Then, the cost of data collection has to do with how many times they ask workers to read a single passage, just to collect a full set of labels.
If chunking the question into smaller subsets (multi-pass), the cost will be greater.
The extreme is when you ask a new worker to read the passage once for every label in the taxonomy.
See Figure \ref{fig:cost} for a breakdown of the cost to carry out this annotation.
}

We see that binary labeling schemes are prohibitively expensive. For this reason we do not include binary labeling schemes in further comparisons. 

Overall, single-pass options are the cheapest (both below $\$2{,}000$), including the cost of training the workers (paying them for completing the tutorial examples). 
Multi-pass options have a higher range, $\$5{,}000-8{,}000$. \textit{Hrchl-pass multi} balances cost (under $\$3{,}000$) but still uses multiple workers to complete one passage's annotations.

\section{Details for producing a proxy for difficulty}
\label{apx:difficulty_details}
\new{Instant agreement (``easy'') are the passage-label pairs for which the entire lab (3 authors plus 3 more students) gave the same value while individually annotating. During this step, the team looked up any terms we were unfamiliar or unclear about, just like the AMT workers are instructed to do.}

\new{Agreement after writing rationales (``medium''): highlight specific parts of the passage and justify why a label should or shouldn’t apply. We only went through this process for passages where we had disagreed in the first step. During this process, we include any level 1 label if there is disagreement within any of its children, even if all team members agreed on the level 1 label.}

\new{For the remaining disagreement, we had a brief (1-2 minute) discussion for every passage-label pair, reading everyone’s given rationale to see whether one of us could convince the others. This process included both reading the rationales written in 1 as well as generating new rationales (in discussion). The passage-label pairs agreed upon in this stage are referred to as ``hard''.}

\begin{figure}[H]
   \centering
   \includegraphics[width=.45\textwidth] {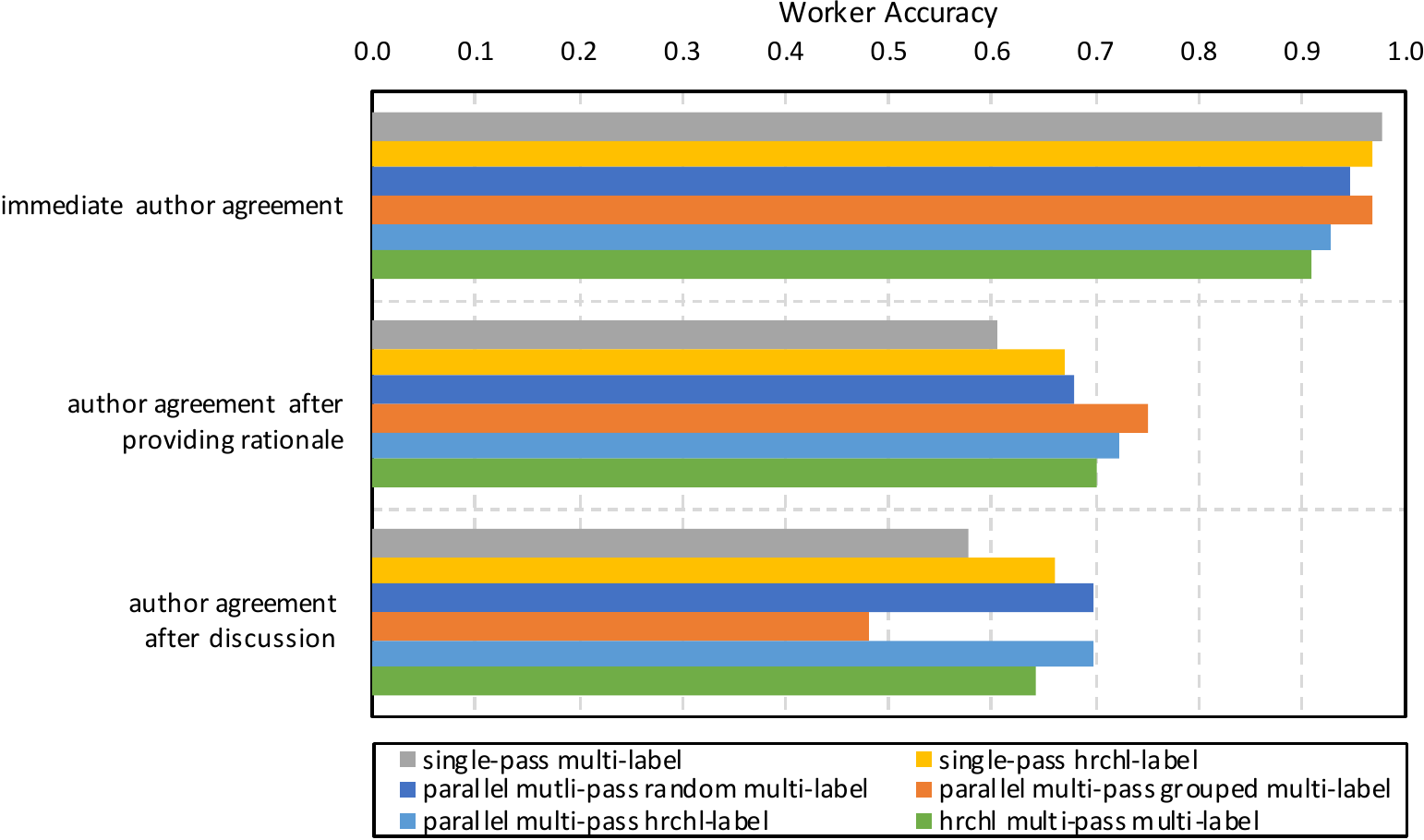}\\
   \vspace{0.2cm}
   \includegraphics[width=.45\textwidth]{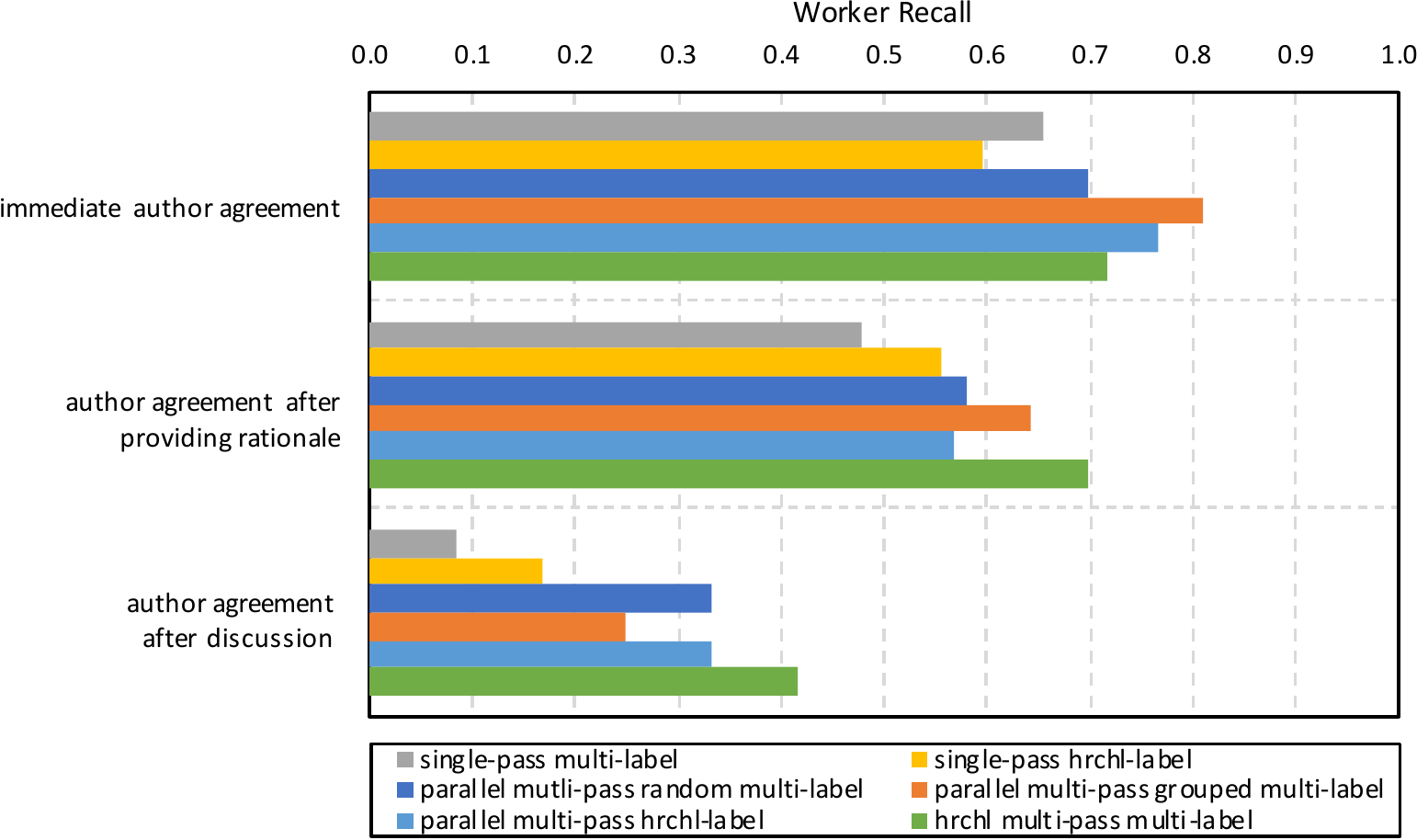}\\
   \vspace{0.2cm}
   \includegraphics[width=.45\textwidth]{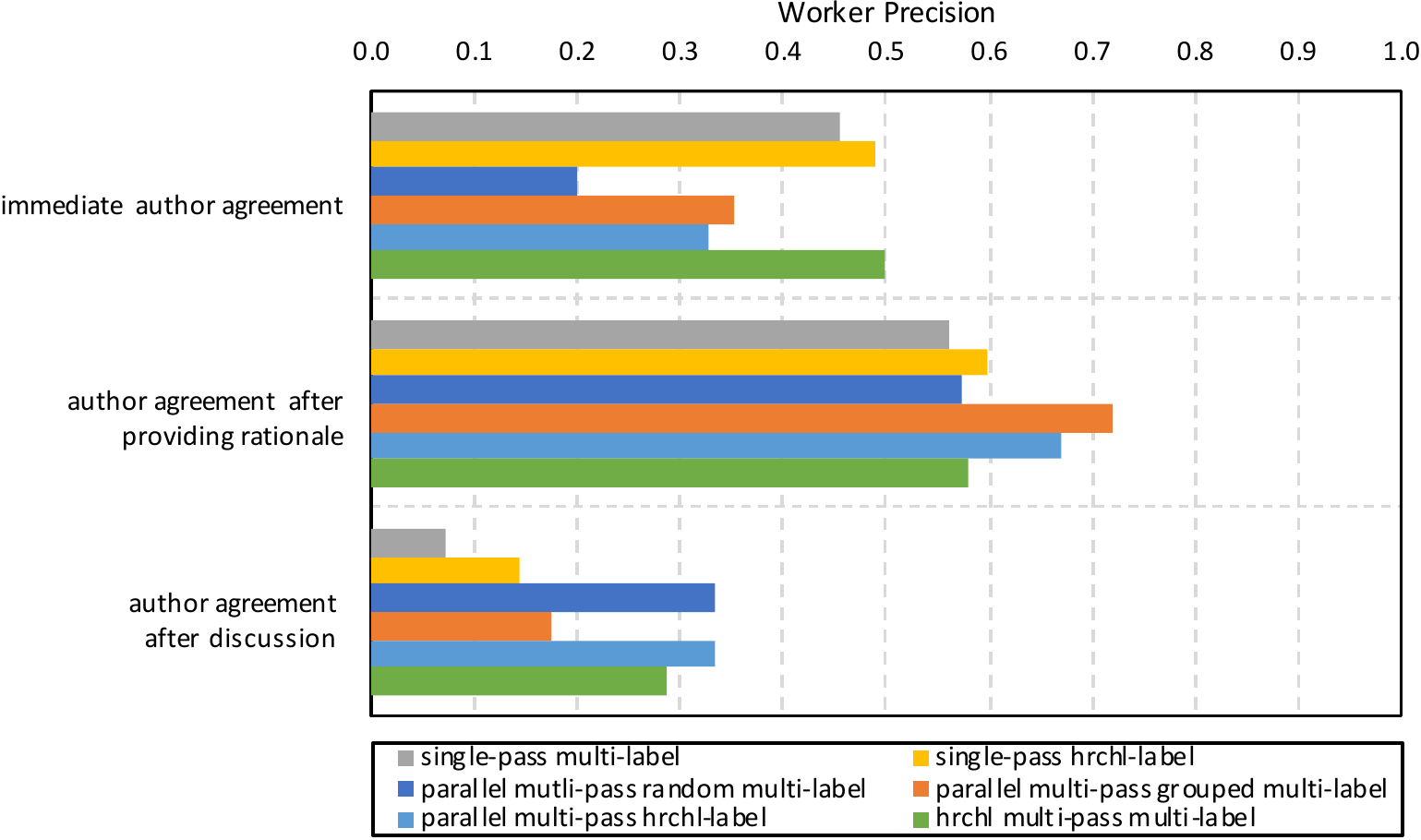}
 \caption{
    \new{
    Worker mean performance metrics for each labeling scheme as difficulty increases (going down). Note that recall decreases steadily for all schemes, while precision largely gets better in the medium category, before experiencing a large dropoff again. 
    Accuracy decreases steadily for most platforms.
    Some platforms have larger drop-offs in performance than others, possibly explaining the difference in F1 score.
    }}
 \label{fig:singlepass_difficulty}
\end{figure}

\new{See Figure \ref{fig:singlepass_difficulty} for a few plots examining the worker performance metrics as each difficulty category varies.}

\new{The remaining passages are marked using each of our individual votes, and placed aside (``no agreement''). We consider these passages to be too subjective to give a gold label for, and don’t evaluate workers on them since any label could be valid so long as they have a strong justification. In future work we may consider looking at the justification/rationale of an AMT worker to assess their performance on highly subjective passage-label pairs.}

\new{We make note of which category each passage-label pair was resolved in, such that we can perform an analysis into how the ``difficulty’’ of passages affect labeling performance. We recognize that this is not necessarily a direct measurement of how difficult it is to label a passage, but we make the assumption that any passage that requires increasingly more thought or discussion to reach consensus will imply this passage is more difficult.}

\section{Performance by taxonomy label}
\begin{figure*}
   \centering
   \includegraphics[width=0.8\textwidth]{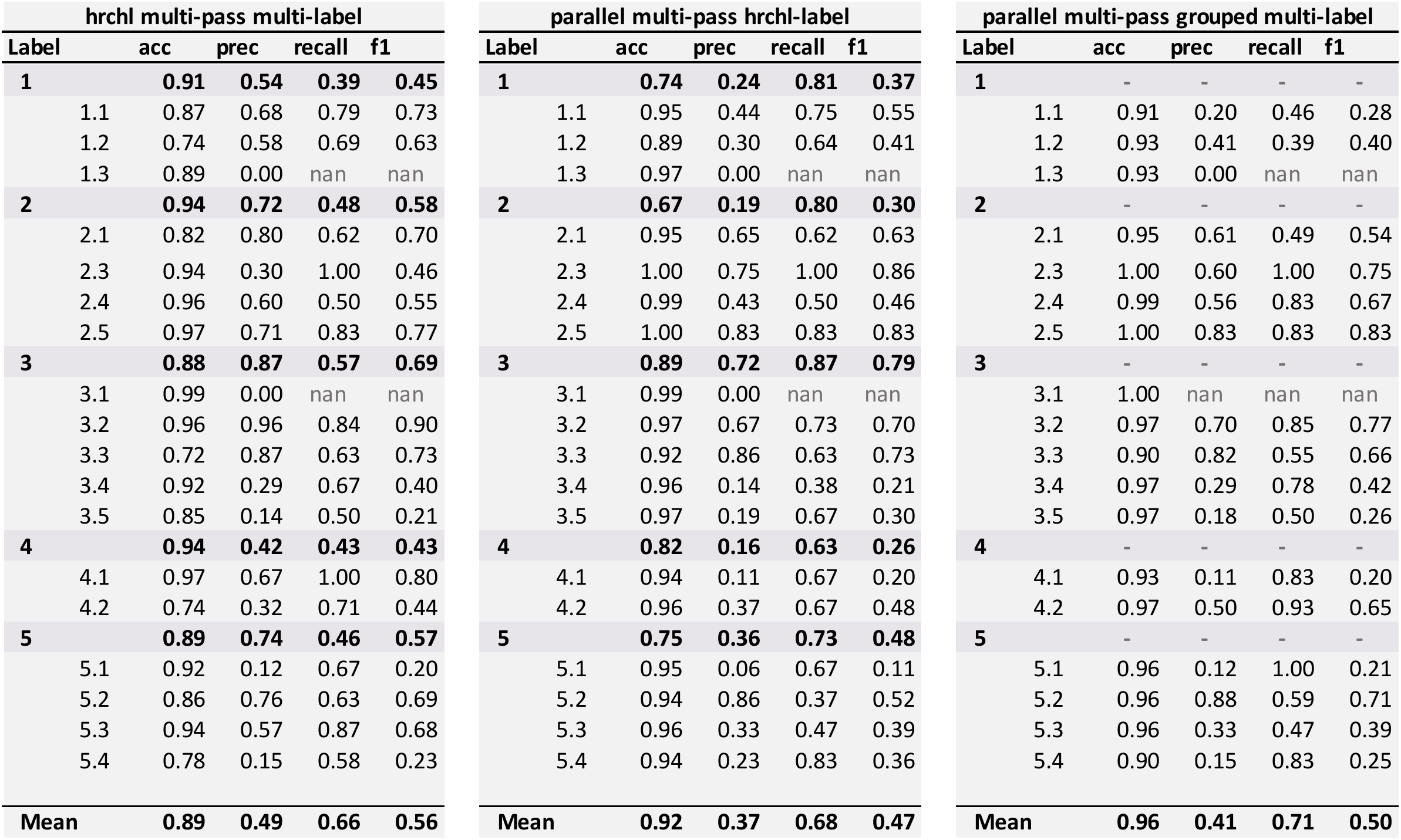} \\
   \vspace{0.2cm}
   \includegraphics[width=0.8\textwidth]{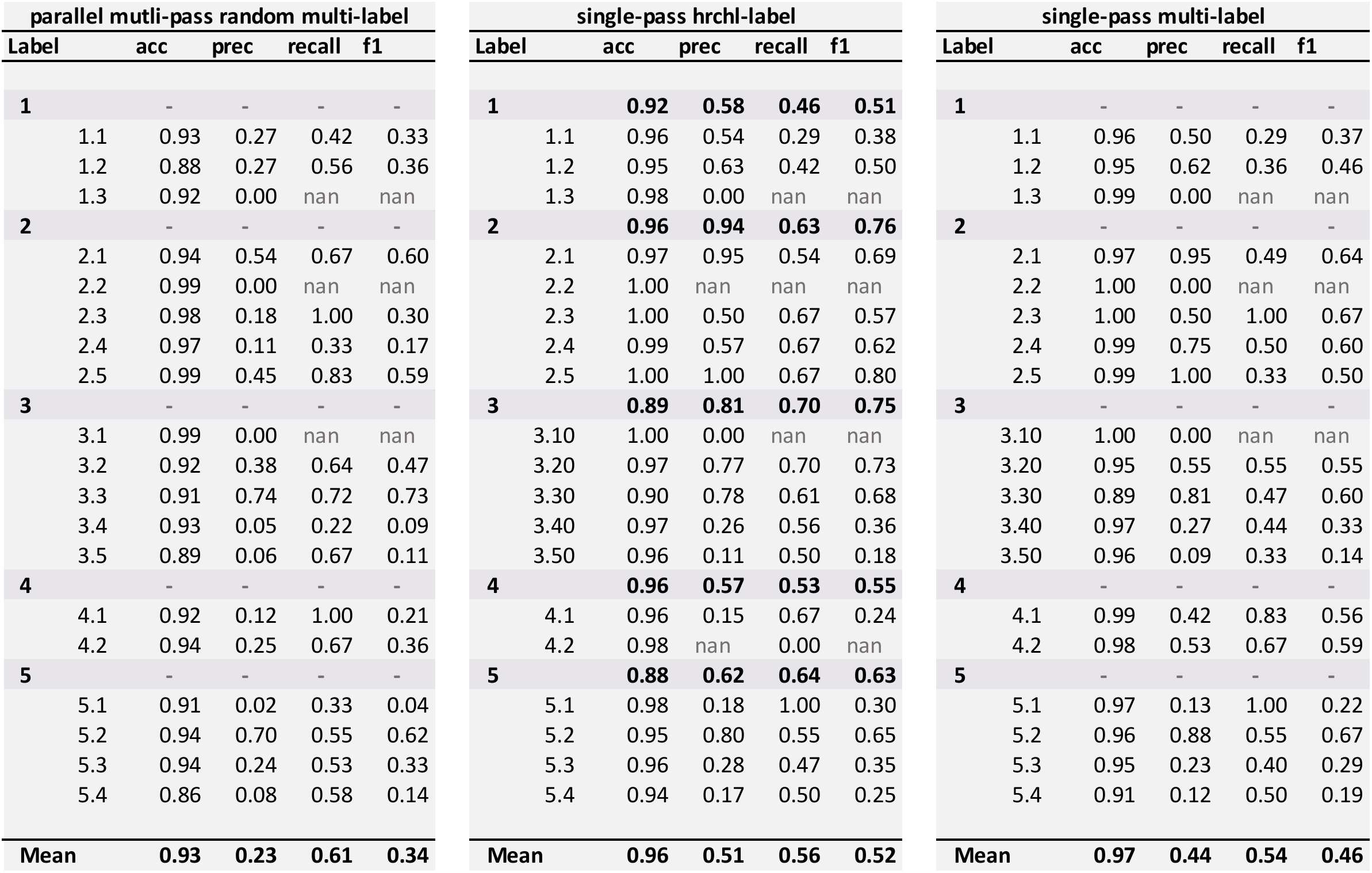}
 \caption{\new{Performance breakdown by taxonomy label for each labeling scheme. Note that nan values appear when there is no positive label given. No positive passages in the ground-truth mean we cannot compute a recall, and no positive passages in the worker annotations mean we cannot compute precision.}}
  \label{fig:perf_by_label}
\end{figure*}

See Tables in Figure \ref{fig:perf_by_label} for each labeling scheme detailing the performance broken down by each individual label of the taxonomy. Final scores are computed for each metric by taking averages across these taxonomy labels.
\label{apx:perf_by_label}

\end{document}